\documentstyle[12pt,epsfig]{ioplppt}

\newcommand {\bbnabla} {\vec{\nabla}}
\newcommand {\dy} {\dot{y}}
\newcommand {\dU} {U^{\prime}}
\newcommand {\ddU} {U^{\prime\prime}}
\newcommand {\dS} {S^{\prime}}
\newcommand {\ddS} {S^{\prime\prime}}
\newcommand {\dr} {\dot{\rho}}
\newcommand {\ddr} {\ddot{\rho}}
\newcommand {\dx} {\dot{x}}
\newcommand {\ddx} {\ddot{x}}
\newcommand {\dg} {\dot{\gamma}}
\newcommand {\dz} {\dot{z}}
\newcommand {\Rmin} {R_{{\rm min}}}
\newcommand {\om}  {\hbar\omega_1}

\begin{document}
\jl{2}

\title{The influence of the dechanneling process on the photon emission
by an ultra-relativistic positron channeling in a periodically bent
crystal.
\footnote{published in
J. Phys. G: Nucl. Part. Phys. {\bf 27} (2001) 95--125, Copyright 2001
IOP Publishing Ltd., http://www.iop.org
}
}

\author{Andrei V. Korol\dag\ddag\ftnote{5}{E-mail:
korol@th.physik.uni-frankfurt.de,\ korol@rpro.ioffe.rssi.ru},
Andrey V. Solov'yov\dag\S\ftnote{6}{E-mail: 
solovyov@th.physik.uni-frankfurt.de,\  solovyov@rpro.ioffe.rssi.ru},
and Walter Greiner\dag\ftnote{7}{E-mail: greiner@th.physik.uni-frankfurt.de}
}

\address{\dag Institut f\"ur Theoretische Physik der Johann Wolfgang
Goethe-Universit\"at, 60054 Frankfurt am Main, Germany}

\address{\ddag Department of Physics,
St.Petersburg State Maritime Technical University,
Leninskii prospect 101, St. Petersburg 198262, Russia}

\address{\S A.F.Ioffe Physical-Technical Institute of the Academy
of Sciences of Russia, Polytechnicheskaya 26, St. Petersburg 194021,
 Russia}


\begin{abstract}
We investigate, both analytically and numerically, the influence of
the dechanneling process on the parameters of undulator radiation
generated by ultra-relativistic positron channelling along a crystal
plane, which is periodically bent.  The bending might be due either to
the propagation of a transverse acoustic wave through the crystal, or
due to the static strain as it occurs in superlattices.  In either
case the periodically bent crystal serves as an undulator which allows
to generate X-ray and $\gamma$-radiation.

We propose the scheme for accurate quantitative treatment of the
radiation in presence of the dechanneling.  The scheme includes (i)
the analytic expression for spectral-angular distribution which
contains, as a parameter, the dechanneling length, (ii) the simulation
procedure of the dechanneling process of a positron in periodically
bent crystals.  Using these we calculate the dechanneling lengths of 5
GeV positrons channeling in Si, Ge and W crystals, and the
spectral-angular and spectral distributions of the undulator over
broad ranges of the photons.  The calculations are performed for
various parameters of the channel bending.
\end{abstract}


\section{Introduction}

\begin{figure}
\hspace{2.5cm}\epsfig{file=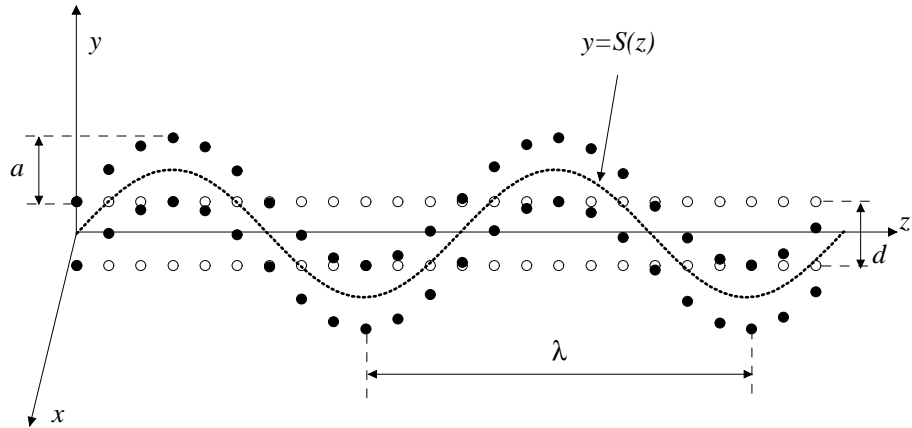,width=12cm, angle=0}
\caption{Schematic representation of the periodically bent 
crystallographic plane.  The open circles mark the atoms belonging to
two neighbouring crystallographic planes (which are parallel to the
$(xz)$ plane) in the initially linear crystal ($d$ is the interplanar
spacing).  The filled circles denote the positions of the atoms when
the crystal is periodically bent.  The $z$ axis is directed along the
straight channel centerline, the $y$ axis marks the direction
perpendicular to the crystallographic planes in the straight crystal.
The profile of the bent channel centerline (the dashed curve) is
described by a periodic function $y=S(z)$ the amplitude, $a$, and the
period, $\lambda$, of which satisfy $a\ll \lambda$.}
\label{fig.1st}
\end{figure}

In this paper we proceed with the investigation of the properties of
the spectrum of emitted photons accompanying ultra-relativistic
positron planar channeling through a crystal which is periodically
bent, as it is illustrated in figure \ref{fig.1st}.  This phenomenon
was described recently in \cite{JPG,laser} and was called Acoustically
Induced Radiation (AIR).  It was noted that the periodic pattern of
crystal bendings (which can be achieved either through propagation of
a transverse acoustic wave or by using static periodically strained
crystalline structures \cite{laser,Uggerhoj2000}) gives rise to a new
mechanism of electromagnetic emission of the undulator type, in
addition to a well-known ordinary channelling radiation
\cite{Kumakhov,Barysh76}.  The mechanism of the AIR is as follows.
Provided certain conditions are fulfilled \cite{laser} the beam of
particles, which enters the crystal at a small incident angle with
respect to the curved crystallographic plane, penetrates through the
crystal following the bendings of its channel.  This results in
transverse oscillations of the beam particles (additional to the
oscillations inside the channel due to the action of the interplanar
force).  These oscillations become an effective source of spontaneous
radiation of undulator type due to the constructive interference of
the photons emitted from similar parts of the trajectory.  It was
demonstrated \cite{laser} that the system ``ultra-relativistic charged
particle + periodically bent crystal'' serves as a new type of
undulator, and, consequently, as a new source of undulator radiation
of high intensity, monochromaticity and of a particular pattern of the
angular-frequency distribution.

The main subject of the present study is the detailed and accurate
quantitative consideration of the influence of the dechanneling
process \cite{Lindhard} on the parameters of the AIR which are the
intensity and the specific pattern of the spectral-angular and
spectral distributions.  The dechanneling, i.e. the decrease in the
volume density of the channeled particles $n(z)$ with penetration
distance $z$ due to the multiple scattering of the projectiles with
the target electrons and nuclei, is the parasitic effect which leads
to the restriction on the crystal length $L$ and, correspondingly, on
the number of the undulator periods $N$, which, in turn, defines the
intensity of the AIR radiation \cite{laser}.  The factor, which plays
a crucial role in obtaining the accurate data for the AIR
characteristics, is a so-called dechanneling length $L_d$, which is a
mean penetration distance covered by a channeling particle.  The
quantity $L_d$ depends on the projectile energy, the type of the
crystal and crystallographic plane, and on the parameters of the
channel bending (these are explained in more detail below in this
section).  In this connection the following problems are solved and
discussed below in the paper:
\begin{itemize}

\item[(i)]
We propose the scheme for accurate quantitative treatment of the AIR
in presence of the dechanneling (section \ref{Yesdech}).  As a result,
we evaluate simple analytic expression for spectral-angular
distribution of the AIR which contains, as a parameter, the
dechanneling length.

\item[(ii)]
We introduce the algorithm of the simulation procedure of the
dechanneling process of a positron in periodically bent crystals
(section \ref{Dechanneling}).

\item[(iii)]
Using (ii) we calculate the dechanneling lengths of 5 GeV positrons
channeling in Si, Ge and W crystals the (110) plane of which are
periodically bent (section \ref{Numerical1}).  The calculations are
performed for various parameters of the channel bending.

\item[(iv)]
Using (i) and (iii) we calculate the spectral-angular and spectral
distributions of the AIR over broad ranges of the photons emitted by 5
GeV positrons channeled in Si and W (section \ref{specrtaAIR}).
\end{itemize}

In our  previous papers \cite{laser} (see also 
\cite{losses,new_lett}) we have estimated the role of the
dechanneling by assuming simple relation $L<L_d$ to be valid.  The
magnitude of $L_d$ was estimated on the basis of the diffusion theory
(see \cite{Kumakhov1973,Waho,Kumakhov1986,Biryukov}) with some
corrections introduced due to the bending of the crystal
\cite{Biryukov,Taratin98}. On the basis of these estimates we carried
out the calculations of the AIR spectra for various energies of the
projectile, different types of the crystals and crystallographic plane
and for various parameters of the channel bending (these are explained
in more detail below).  It was also pointed out \cite{laser} that the
system ``ultra-relativistic charged particles + periodically bent
crystal'' leads, in addition to the spontaneous radiation, to the
possibility of generating stimulated emission, similar to the one
known for a free-electron laser \cite{Madey71} in which the
periodicity of a trajectory of an ultra-relativistic projectile is
achieved by applying a spatially periodic magnetic field.  In
connection with the stimulated AIR it was noted in \cite{laser} that
to achieve a noticeable degree of amplification one has to operate
with a positron bunch of a high volume density.  The dechanneling
length played a crucial role in these estimations by influencing the
intensity of the spontaneous AIR as well as the gain factor and the
corresponding value of the positron bunch volume density for the
stimulated AIR.  Indeed, increasing of the dechanneling length by a
factor of 2 results in decreasing of the channeling beam density
needed to achieve large gain almost by by a factor of 10 \cite{laser}.

By utilizing the numeric scheme outlined in section \ref{Dechanneling}
(to our knowledge it is the first time that the dechanneling process
is simulated in a periodically bent crystal) we demonstrate further in
section \ref{Numerical1} that the simple estimation of $L_d$ made in
\cite{laser} in many cases produces the results which are noticeably
smaller then more accurate ones.  This results, consequently, in
essential (up to the order of magnitude for heavy crystals) difference
in the intensities of the AIR radiation (section \ref{specrtaAIR})
computed by using the estimated, $L_d^e$, and the calculated, $L_d^c$,
values of the dechanneling lengths.

Prior to presenting our main results let us formulate the important
criteria which are used further in the paper.

The planar channeling of an ultra-relativistic projectile in a
straight crystal (for reviews see \cite{Gemmell,Baier,KumakKomar})
occurs only if the entrance angle of the projectile velocity with
respect to the mid-plane does not exceed the critical angle of
channeling
\cite{Lindhard}  
\begin{equation}
\Theta < \Theta_L 
\label{cond0}
\end{equation}
where the quantity $\Theta_L$ (often called as the Lindhard angle) is
defined as $\left(2 U_{\rm max}/m \gamma\right)$.  Here $U_{\rm max}$
is the interplanar potential-well depth, $\gamma$ is the relativistic
factor of the projectile and $m$ is its mass (below we consider only
the channeling of positrons, if otherwise is not explicitly stated).

The criterion for a stable channeling of an ultra-relativistic
particle in a bent crystal was formulated in \cite{Tsyganov} and has
clear physical meaning: the maximum centrifugal force due to the
channel bending must be less than that of the interplanar field.
Provided this condition is fulfilled, the beam of channeling particles
at each instant moves inside the channel, especially parallel to the
bent crystal midplane as it does in the case of linear channeling.
The dynamics of the particles channeled in crystals bent with a
constant (or slowly varied) curvature $1/R$ is described in details in
\cite{Biryukov,Taratin98,Ellison1982,RelCha,Solovyov,NIMB2000} 
(see also the references therein). In the cited reviews the main
accent was made on channeling of heavy projectiles due to the growing
interest for using bent crystals for extraction, bending and splitting
of high-energy beams (for the latest review see \cite{NIMB2000} and
the references therein.

For a periodically bent crystal the criterion for a stable channeling
can be written in the form \cite{laser}
\begin{eqnarray}
C = { \varepsilon \over R_{{\rm min}}\, \dU_{{\rm max}}} 
\ll 1,
\label{cond1}
\end{eqnarray}
where $\dU_{{\rm max}}$ is the maximum value of the interplanar field,
$R_{{\rm min}}$ is the minimum curvature radius of the bent channel
centerline.  Introducing the shape function $S(z)$, which describes
the profile of the centerline (see figure \ref{fig.1st}), the minimum
curvature $1/R_{{\rm min}}$ can be written in terms of the amplitude
$a$ and the period $\lambda$ of $S(z)$: $1/R_{{\rm min}}\approx
|\ddS(z)|_{{\rm max}}^{-1} \sim a^{-1}(\lambda/2\pi)^2$.  Thus, for a
given crystal and its crystallographic plane, the condition
\ref{cond1} establishes the ranges of projectile energies and the
parameters $a$ and $\lambda$ inside which the channeling in a
periodically bent channel can effectively occur.

The channeling radiation \cite{Kumakhov,Barysh76} formed by
ultra-relativistic positrons channeled in various straight crystals
has been studied extensively both theoretically
(e.g. \cite{KumakKomar,Baier,BaryshKniga,Bazylev} and references
therein) and experimentally
\cite{Avakian82,Uggerhoj1983,BakI,Uggerhoj1993,Dulman}.

Several papers \cite{Taratin,Arutyunov,Solovyov} have treated the
radiation emitted by light projectiles channeled in non-periodically
one-arc bent crystals.  In this case the channel bending gives rise to
the synchrotron-type radiation in addition to the channeling one.  In
the cited papers it was demonstrated that for small values of the
parameter $C$ (i.e. for large curvatures of the bending) the photon
energies corresponding to the maximum intensity of the synchrotron
radiation do not exceed the energy of the first harmonic of the
channeling radiation.

When the shape of the bent channel becomes periodic the 
additional radiation becomes of the undulator type 
\cite{Ginz,Baier,Bazylev,Barbini} rather than the synchrotron one.
Then, there appear two essentially different regimes which are defined
by the magnitude of the ratio $a/d$ \cite{laser,new_lett}.  In the
case $a/d \ll 1$, which can be realized either by applying
low-amplitude transverse ultrasonic wave
\cite{Ikezi84,Ketterson86,Armyane86,BaryshJPC90,Dedkov94} 
or by using a supperlattice crystalline structure \cite{Ketterson86},
the characteristic frequencies of the two types of radiation can
become compatible, and thus, one can consider, at least theoretically,
the phenomenon of the resonant enhancement of the photon yield due to
the coupling of these mechanisms.  However, as it was demonstrated in
\cite{laser,losses,new_lett}, in the limit $a/d \ll 1$ the
intensity of AIR is noticeably small compared not only with that of
the channeling radiation but also with the intensities of the
radiative background mechanisms, such as coherent and incoherent
bremsstrahlung.  Hence, it is highly questionable whether the AIR can
be considered as a new phenomenon in the limit of low values of $a$.

On the contrary, if the amplitude of the shape function and the
interplanar spacing satisfy strong inequality
\begin{eqnarray}
a\gg d
\label{cond2}
\end{eqnarray}
then (i) the characteristic frequencies of the AIR and the channeling
radiation are well-separated (this is the consequence of {\it both}
conditions, (\ref{cond2}) and $C\ll 1$), (ii) the intensity of the AIR
is essentially higher that that of the ordinary channeling radiation
\cite{laser,losses,new_lett}.  In this case the
undulator-type radiation due to the periodic structure of the crystal
bending can be considered as a new source of the emission within the
$X$- and $\gamma$-range.

Below in the paper we assume that both strong inequalities
(\ref{cond1}) and (\ref{cond2}) are fulfilled.  It is important to
mention that in our scheme the amplitude $a$ is subject not only to
the condition (\ref{cond2}) but also is much smaller compared with the
period of $S(z)$: $a \ll \lambda$.  Hence, neither the interplanar
spacing nor the distance between neighbouring lattice atoms are
changed (in more detail it is discussed in section \ref{EM}).

To conclude the introductory part we note that two realistic ways of
``preparation'' of a periodically bent channel may be discussed
\cite{JPG,laser}.  It is feasible, by means of modern technology, to
grow the crystal with its channels been statically bent according to a
particular pattern. In particular, the crystalline undulator can be
constructed based on graded composition strained layers in a
superlattice \cite{Uggerhoj2000} (see also \cite{Breese,Nakayama}).
Another possibility is to use high-amplitude transverse ultrasonic
wave.  In the latter case the parameters of the undulator can be
easily tuned by varying the amplitude and the frequency of the
acoustic wave.  It was demonstrated \cite{laser} that the
acoustically-based undulator can be created by applying the supersonic
waves within the frequency range $10\dots 100$ MHz which is achievable
in the experiments on propagating the positron beams through
acoustically-excited crystals \cite{Avakian98}.

\section{The AIR radiation in  absence of the dechanneling}
\label{Nodech}

Let us consider the spectral-angular distribution of the photons
emitted during the projectile positron planar channeling in a
periodically bent crystal along the $z$ direction 
(see figure \ref{fig.1st}). 
The motion of the particle occurs in the   $(yz)$-plane, so that its 
radius vector and the velocity are given by
${\bf r}(t)= z\,{\bf e}_z +y\,{\bf e}_y $,
${\bf v}(t)= \dot{z}\,{\bf e}_z +\dot{y}\,{\bf e}_y $ where
${\bf e}_z$ and ${\bf e}_y$  are the unit vectors along the $z$- and
$y$-directions, respectively.

Provided the conditions (\ref{cond0}-\ref{cond2}) are fulfilled
the photon spectrum is formed due to the
ordinary channeling mechanism and to the AIR, which
prevail over two other radiative background mechanisms, 
coherent and incoherent bremsstrahlung. 
It has been established, both theoretically and experimentally 
(e.g. \cite{Baier,Uggerhoj1983,Uggerhoj1993}), that 
the total radiative losses of a positron  channeling in 
a straight crystal are essentially smaller than its energy  
provided  $\varepsilon \leq 10-20$ GeV. 
This estimate is correct in the case of channeling through periodically  
 bent crystal  as well \cite{losses}.

Therefore, restricting from above the range of $\varepsilon$  by 
the value $\sim 10 GeV$, one can assume that the emitted photon
energies satisfy $\hbar \omega \ll \varepsilon$. 
This inequality allows to consider the process of the photon emission
within the framework of classical electrodynamics (e.g. \cite{Land2}).

The distribution of the energy radiated in the cone
$\d \Omega_{\bf n}$ along the direction ${\bf n}$ 
and within the frequency interval
$\left[\omega, \omega+\d\omega\right]$ 
is given by the following expression
\begin{equation}
{ \d E \over \d \omega \d \Omega_{\bf n}} =
\hbar\,\alpha \, {\omega^2  \over 4 \pi^2 } \, 
\int_0^{\tau}\, \int_0^{\tau}\, \d t_1\,\d t_2 \
\e^{\i \omega \phi(t_1,t_2)}
\ f(t_1,t_2)
\label{eq.1}
\end{equation}
where $\alpha= e^2/ \hbar\, c \approx 1/137$,
$\tau$ is the time of flight through the crystal of thickness $L$, 
$\tau \approx  L/c$.
The solid angle $\Omega_{\bf n}$ of the photon emission is
characterized by two polar angles $\vartheta$ and $\varphi$.
The functions $\phi(t_1,t_2)$ and 
$f(t_1,t_2)$ are given by
\begin{equation}
  \phi(t_1,t_2) = t_1 - t_2 - {1 \over c}\,
{\bf n}\cdot ({\bf r}_1 - {\bf r}_2),
\qquad
 f(t_1, t_2)= {1 \over 2}
\left(
{ {\bf v}_1\cdot {\bf v}_2 \over c^2} -1
\right)
\label{wkb.2}
\end{equation}
The short-hand notations used here are ${\bf r}_{1,2}= {\bf r}(t_{1,2})$
and ${\bf v}_{1,2}= {\bf v}(t_{1,2})$.

Assuming that the relativistic factor of the projectile satisfies
strong inequality $\gamma \gg 1$, one can carry out the expansion of
the functions $\phi(t_1,t_2)$ and 
$f(t_1,t_2)$ in powers of $\gamma^{-1}$ (see e.g. \cite{Baier}).
Then, retaining the senior non-vanishing terms, one transforms the r.h.s. of 
(\ref{eq.1}) to a form which is more suitable for
analytical and numerical analysis:
\begin{equation}
{\d E \over \d\omega\,\d\Omega_{\bf n}} 
=
\hbar\,\alpha \,
{\omega^2 \over 4\pi^2}\, 
\left\{
\right|I_1 - \vartheta\, \cos\varphi\, I_0\left|^2
+
\vartheta^2\, \sin^2\varphi
 \right|I_0\left|^2
\right\}
\label{1_1.1}
\end{equation}
where
\begin{equation}
I_0
=
\int_0^{\tau}\, \d t\,
\e^{\i \omega \tilde{\phi}(t)},
\qquad
I_1
=
\int_0^{\tau}\, \d t\,
{\dy(t) \over c}\,
\e^{\i \omega \tilde{\phi}(t)}
\label{1.1.2}
\end{equation}
The modified phase function $\tilde{\phi}(t)$ reads as
\begin{eqnarray}
\tilde{\phi}(t)
=
t\, \left({1 \over 2 \gamma^2} + {\vartheta^2 \over 2} + 
{p^2 \over 4 \gamma^2} \right)
+
{\Delta(t) \over 2}
-
\vartheta\, \cos\varphi\, {y(t) \over c}
\label{1.1.9a}\\
\Delta(t)
=
\int^t \d t^{\prime}
\left(
{\dy^2(t^{\prime}) \over c^2} 
-
{p^2 \over 2\gamma^2}
\right),
\label{1.1.9b}\\
p^2 = 2\gamma^2 \,{\overline{\dy^2} \over c^2} 
\label{p2}
\end{eqnarray}
The last relation defines the so-called undulator parameter $p$, 
which is commonly used in the theory of the undulator
radiation (e.g. \cite{Baier}).
This quantity is related to 
the mean-square velocity  $\overline{\dy^2}$ 
in the $y$-direction, which is transverse with respect to the
centerline of the bent channel (see figure \ref{fig.1st}). 

If one neglects the processes of inelastic scattering by the 
crystal electrons and nuclei, then the motion of the 
positron in the channeling regime can be described within 
the framework of the continuum approximation \cite{Lindhard,Gemmell} 
for the interaction potential $U$ between the projectile and lattice
atoms.
Within this approximation and in the case of
periodically bent channel the dependence
$y(t)$, which enters the r.h.s. of (\ref{1.1.2}-\ref{1.1.9b}) and, thus,
defines the  spectral-angular distribution (\ref{1_1.1}), can be
obtained as follows \cite{losses,new_lett}.

At any given moment $t$ the $y$-coordinate of the projectile 
can be presented as a sum of two terms
\begin{eqnarray}
y(t)
=
S(z)
+
\rho(t)
\label{y_t}
\end{eqnarray}
Here $S(z)$ is the displacement due to the channel bending. 
The quantity $\rho(t)$ stands for the displacement of the projectile
from the centerline of the channel. 
The equation of motion for $\rho(t)$ reads \cite{Ell_Pic81}
\begin{eqnarray}
\ddot{\rho}
=
-{U^{\prime} \over m\gamma}
-
c^2\, S^{\prime\prime}
\label{rho_t}
\end{eqnarray}
where the dot sign stands for the time derivative, 
$U^{\prime}=\d U/\d \rho$ is the derivative of the interplanar
potential which depends on $\rho$, and
$S^{\prime\prime}= \d^2 S/ \d z^2$.
The first term on the r.h.s. represents the acceleration of the particle
due to the action of the interplanar force, while the second 
one is the acceleration due to the centrifugal force.
Note that on the r.h.s. one may put 
$z=c t$ after carrying out the differentiation $S^{\prime\prime}$,
because all the deviations from $c$ of the particle's velocity 
along the $z$-axis were taken into account when writing
eqs.  (\ref{1.1.2}-\ref{1.1.9b}). 

Solving the equation of motion (\ref{rho_t}) and, then, using the
result, combined with (\ref{y_t}), in (\ref{1.1.2}-\ref{1.1.9b}), one
obtains the total spectral-angular distribution of the radiation
(\ref{1_1.1}).
This scheme was used in our recent papers \cite{new_lett}
where the results of the numerical calculations of the spectra were
presented for various shapes $S(z)$ and with using different
approximations for the interplanar potential $U(\rho)$.
 
A particle which channels through periodically bent crystal
experiences two types of oscillations in the transverse direction, 
as it is seen from (\ref{y_t}).
Firstly, there are the channeling oscillations due to the action of 
the interplanar potential. 
The oscillations of the second type (henceforth we refer to them as
to the ``undulator oscillations'') are related to the periodicity 
in the shape $S(z)$ of the centerline of the channel inside which the
particle moves.
The characteristic frequencies of these oscillations, denoted as 
$\Omega_{ch}$ and $\Omega_{u}$ respectively, can be estimated as
follows
\begin{equation}
\Omega_{ch} 
\sim 
\sqrt{{U^{\prime} \over  d\, m \gamma}},
\qquad
\Omega_{u} 
=
{2\pi c \over \lambda}
\label{1.1.13a}
\end{equation}
Correspondingly, the quantities 
$\Omega_{ch}, \Omega_{u}$ define the characteristic 
frequencies, $\omega_{ch}$ and $\omega_{u}$, of the photons emitted 
due to the channeling and the undulator oscillations:
\begin{equation}
\omega_{ch} 
\sim 
\gamma^2 \,\Omega_{ch},
\qquad
\omega_{u} 
\sim 
\gamma^2 \,\Omega_{u}
\label{1.1.14}
\end{equation}
Taking into account the relations (\ref{cond1}) and (\ref{cond2}) 
one estimates the relative magnitude of $\Omega_{ch}$ and
$\Omega_{u}$:
\begin{equation}
{\Omega_{u}^2 \over \Omega_{ch}^2} 
\sim 
{\omega_{u}^2 \over \omega_{ch}^2} 
\sim
{d \over a}\, {\varepsilon \over  R_{{\rm min}}\,\dU_{{\rm max}}} 
=
{d \over a}\, C \ll 1
\label{cond3}
\end{equation}
Hence, the characteristic frequencies of the AIR  and the ordinary 
channelling  radiation are well separated \cite{losses}, and,
what is also very important, in the region  $\omega \sim \omega_{u}$
there is no coupling of the 
two mechanisms of the radiation \cite{new_lett}. 

The results of numerical calculation of the spectral distribution 
(\ref{1_1.1}) presented in \cite{new_lett}
clearly suggest that the characteristics of the AIR spectra 
(the position of the peaks of the radiation, their width, the
radiated intensity) are, practically, 
insensitive to the choice of the approximation used 
to describe  the interplanar potential $U(\rho)$. 
Moreover, if being interested in the spectral-angular 
distribution only in the region $\omega \sim \omega_{u}$, then 
one may disregard the channeling oscillations and to assume that the
projectile moves along the centerline of the bent channel.
In this case, taking into account the periodicity of the undulator
motion, the r.h.s. of (\ref{1_1.1}) can be presented in the  form 
which clearly exhibits all the characteristic features  
of the AIR radiation.

Let assume that there is integer number of 
the undulator periods on the crystal length,
$L = \lambda\, N$. 
Also, for the sake of transparency of the analytic expressions, let
consider the shape function $S(z)$ of the form
\begin{equation}
S(z) =
a\, \sin kz,
\qquad
k=2\pi/\lambda
\label{sinSz}
\end{equation}
Let assume that there is integer number of 
the undulator periods on the crystal length,
$L = \lambda\, N$.
Then, the spectral-angular distribution of the AIR radiation can be
represented  in the form \cite{laser,Bazylev}
\begin{equation}
{\d E_N \over \d\omega\,\d\Omega_{\bf n}} 
=
\hbar\,\alpha \,
{\omega^2 \over \omega_0^2}\, 
{D_N(\eta) \over 4\pi^2}\,
\left\{
\left|{p \over \gamma} F_1 - \vartheta\, \cos\varphi\, F_0\right|^2
+
\vartheta^2\, \sin^2\varphi
 \left|F_0\right|^2
\right\}
\label{1.2}
\end{equation}
Here the subscript $N$ in the notation $\d E_N$ indicate the number 
of undulator periods, $\omega_0=2\pi c/\lambda$ is the so-called
undulator frequency,
and the quantities $F_m$ ($m=0,1$) stand for the integrals
\begin{eqnarray}
\fl
& F_m
=
\int_0^{2\pi}
\d \psi\,
\cos^m\psi \,
\exp\left(
\i \left[\eta \psi + {x \over 2}\sin(2\psi) - \beta \sin \psi\right]
\right),
\label{1.3a}\\
\fl
&
\eta = {\omega \over \omega_0}\,
\left({1 \over 2 \gamma^2} + {\vartheta^2 \over 2 } 
+
{p^2 \over 4\gamma^2}
\right),
\quad
x={\omega \over \omega_0}\,{p^2 \over 4\gamma^2},
\quad
\beta = \vartheta\, \cos\varphi\, 
{\omega \over \omega_0}\,{p \over \gamma}. 
\label{1.3b}
\end{eqnarray}

The information on the specific pattern of the undulator radiation,
namely on the positions of the narrow maxima in the spectral-angular 
distribution (\ref{1.2}), is accumulated in the factor
$D_N(\eta)$ which equals to
\begin{equation}
D_N(\eta) = 
\left({\sin N\pi \eta \over \sin \pi \eta}  
\right)^2
\label{1.4}
\end{equation}
This function, which is well-known in the theory of undulator
radiation (see e.g. \cite{Baier,Bazylev}) and in classical
diffraction theory \cite{Land2}, 
 defines the profile of the line of the emission in an
ideal undulator. 
It has sharp maxima at the points $\eta = k =
1,2,3...$ where $D_N(k)= N^2$. 
The integer values $\eta = k$ define 
(see the first equation in (\ref{1.3b})) 
the characteristic frequencies $\omega_k$
(harmonics) of the undulator radiation for which the spectral-angular
distribution (\ref{1.2}) reaches its maxima:
\begin{equation}
\omega_k = 
{4\gamma^2\, \omega_0\, k \over 2 + 2\,\gamma^2 \vartheta^2 + p^2}
\label{1.4a}
\end{equation}
The natural width of each line of the emission can be estimated as
$\Delta\omega^{\prime} = (1/N) (\omega_k/k)$ and is independent 
on the harmonic number $k$.

The largest number of the radiated harmonics, $k_{{\rm max}}$, can be
estimated as $k_{{\rm max}}\sim p^3/2$ 
(see e.g. \cite{Baier,Bazylev}), so
that in the range $\omega \leq \omega_{k_{{\rm max}}}$ the spectral-angular
distribution of the AIR represents by itself the set of narrow
well-separated peaks. The radiation intensity at $\omega =\omega_k $
is proportional to the square of the total number of the undulator periods, 
$N^2$, reflecting the coherence effect of radiation.

It follows from (\ref{1.4a}) that the position of the peaks 
are dependent on the relativistic factor, on the emission angle 
$\vartheta$, and on the undulator parameter $p$.

The only feature, related to the channeling oscillations, 
which modifies the formulae (\ref{1.2}-\ref{1.4a}), which 
describe the AIR spectrum, 
concerns the definition of the undulator parameter 
$p$ (see (\ref{p2})).

When the channeling oscillations are disregarded completely 
(this means omitting the term $\rho(t)$ on the r.h.s. of 
(\ref{y_t})) then the undulator parameter equals 
$2\gamma^2 \overline{S^{\prime\, 2}}=4\pi^2\gamma^2 a^2/\lambda^2$, 
and for given $\gamma$ is defined only by the
amplitude and the period of the shape function $S(z)$.

To account for the channeling oscillations and to modify the
definition of $p$ one can use the following arguments.
The quantity $p^2$ is proportional to the 
the mean-square velocity  $\overline{\dy^2}$. 
The latter, according to (\ref{y_t}), can be written as follows:
\begin{eqnarray*}
\overline{\dy^2}
=
c^2\,\overline{S^{\prime\, 2}}
+
\overline{\dot{\rho}^2}
+
2 c\,\overline{S^{\prime}\,\dot{\rho}}
\end{eqnarray*}
The last term on the r.h.s. contains the product of the factors
oscillating with incompatible frequencies (see (\ref{cond3}), and,
therefore, it disappears when one carries out the averaging over the
time interval $\sim 1/\Omega_{ch}$.
Hence, the square of the undulator parameter 
can be  written as follows
\begin{eqnarray}
p^2
=
\gamma^2\, \left({2\pi\, a \over \lambda}\right)^2
+
\gamma^2\, 
\overline{\dot{\rho}^2}
\label{p2_mod}
\end{eqnarray}
The results presented in \cite{new_lett} demonstrate that 
in the region $\omega\sim \omega_u$
the spectral-angular distribution, 
obtained by means  of the exact procedure which includes
(i) numerical integration of the equation of motion 
(\ref{rho_t}) with the interplanar potential considered within the 
the Moli\`ere approximation, 
(ii) subsequent evaluation of the integrals in 
(\ref{1_1.1}), and 
(iii) by further averaging of the result over all channeling 
trajectories,  
can be accurately reproduced by using (\ref{1.3b}) and 
(\ref{p2_mod}) with the term $\overline{\dot{\rho}^2}$ 
calculated within the harmonic approximation for $U(\rho)$.
The use of (\ref{1.3b}) instead of (\ref{1_1.1})
simplifies the numerical procedures considerably, 
leading to the reduction, by orders of magnitude, of the 
CPU time.

\section{The AIR radiation in presence of the dechanneling}
\label{Yesdech}

If the  dechanneling is neglected, one may unrestrictedly
increase the intensity of the AIR by considering larger
$N$-values. 
In reality, random scattering of the channeling 
positron by the electrons and nuclei of the crystal leads 
to a gradual increase of the particle energy
associated with the transverse oscillations in the channel. 
As a result, the transverse energy at some distance from the
entrance point exceeds the depth of the interplanar potential well,
and the particle leaves the channel. 
Consequently, the volume density $n(z)$ of the channeled particles
decreases with the penetration distance $z$. 
Although the exact explicit dependence of the channeled fraction of the
particles $n(z)/n_0$ on $z$ 
($n_0$ is the volume density of the channeled particles
at the entrance) hardly can be obtained by analytical means due 
to the complexity of the accurate treatment of the multiple scattering 
problem in medium, it has been argued \cite{Biryukov,KumakKomar}
that far from the entrance point $n(z)/n_0$ 
can be described by the exponential decay law
\begin{equation}
{n(z) \over n_0} = \exp\left(-  z/ L_d\right)
\label{1.5}
\end{equation}
Basing on this relation, below in this section we present a model which 
allows to modify the spectral-angular distribution (\ref{1.2})
by taking into account the influence of the dechanneling effect. 

Let us assume that the period $\lambda$ of the shape function $S(z)$
and the dechanneling length $L_d$ satisfy the
strong inequality $\lambda \ll L_d$.
If so, then the number of particles (per unit volume)
$\Delta n=n(z)- n(z+\lambda)$, which dechannel 
within the interval $[z, z+\lambda]$, is small compared with the number 
of the particles $n(z)$ which reach the point $z$. 
The inequality allows also to introduce the (discrete) probability $p_j$
of the event, that after channeling through $j$ periods of the
undulator ($j=1,2,\dots N-1$) the particle dechannels within the
subsequent $(j+1)$-th period. 
In accordance with (\ref{1.5}) the probability $p_j$ is written as follows
\begin{equation}
p_j
=
\cases
{
\exp\left(-j/N_d\right)\,
\left(\exp\left(1/N_d\right)-1\right),
& for $j = 1,2\dots N-1$\\
\exp\left(-(N-1)/N_d\right)\,
& for $j = N$\\
}
\label{1.7}
\end{equation}
and is subject to the normalization condition 
$\sum_{j=1}^{N} p_j = 1$.
In (\ref{1.7}) we introduced the quantity $N_d = L_d/\lambda$ which is 
the number of the undulator periods within the dechanneling length.
Since $N_d \gg 1$, then 
the deviation of $N_d$ from the nearest integer 
is of no importance, and henceforce
the quantity $N_d$ is treated as the integer much larger than one. 

With the help of (\ref{1.7}) one can construct the
following quantity, which characterizes the spectral-angular
distribution of the AIR radiation formed in the crystal of the length
$L$ (and of the corresponding number of the undulator periods $N$) 
and which accounts for the dechanneling effect:
\begin{eqnarray}
\left\langle 
{\d E_N \over \d\omega\,\d\Omega_{\bf n}}\right\rangle 
&=
\sum_{j=1}^N \,
p_j\,
{\d E_j \over \d\omega\,\d\Omega_{\bf n}}
\label{1.8a} \\
&=
\hbar \alpha \,
{\omega^2 \over \omega_0^2}\, 
{\left\langle D_N(\eta)\right\rangle  \over 4\pi^2}\,
\left[
|{p \over \gamma} F_1 - \vartheta\, \cos\varphi\, F_0|^2
+
\vartheta^2\, \sin^2\varphi
|F_0|^2
\right]
\label{1.8}
\end{eqnarray}
with
\begin{equation}
\fl
\left\langle D_N(\eta) \right\rangle 
=
\left(\exp\left(1/N_d\right)-1\right)
\sum_{n=1}^N \,
\exp\left(-{n \over N_d}\right)
\,
\left({\sin n \pi \eta \over \sin \pi \eta}  \right)^2
+
\exp\left(-{ N\over N_d}\right)
\left({\sin N \pi \eta \over \sin \pi \eta}  \right)^2
\label{1.9}
\end{equation}

Let us note here the difference between the averaging procedure
suggested above by relation (\ref{1.8a}) and the conventional
scheme (see e.g. \cite{Baier}) for obtaining the characteristics of 
the electromagnetic radiation 
accompanying the motion of an ultra-relativistic particle in medium.
The latter approach prescribes that, in the presence of multiple
scattering,  the radiative spectrum (\ref{eq.1})
formed during the motion of a projectile in an external field
along some particular trajectory must be averaged over all allowed
trajectories. 
This operation is carried out with the help of the
distribution function satisfying the corresponding kinetic equation
\cite{Baier}.
Although the AIR radiation in medium is not an exclusion from this
rule, in this case there is one peculiarity 
which justifies the use of the averaging
procedure (\ref{1.8a}). 
As mentioned above,  the motion of a positron 
channelled through a periodically bent crystal 
is characterized by two essentially different modes. 
The ``fast'' mode corresponds to the channeling oscillations inside the
channel, the ``slow'' mode describes the motion of the particle along
the channel centerline. 
The AIR is due to the ``slow'' mode and in the case $a\gg d$
its characteristics are almost independent on the parameters of the 
fast channeling oscillations \cite{laser,new_lett}.
Hence, if one is interested in the spectral-angular distribution of the 
AIR only, then it is sufficient to totally disregard the channeling
oscillations and to assume that the particle moves along the fixed
trajectory which is defined by the profile of the bent channel
centerline.
The intensity of the AIR radiation is then defined by the number $n$ of
the $\lambda$-periods which the particle passes before leaving the
channel due to the multiple scattering.
The value of $n$ is subject to the probability distribution
(\ref{1.7}) which, in turn, is based on the exponential decay law
(\ref{1.5}).

Hence, to obtain the parameters of the AIR radiation in presence
of the dechanneling it is sufficient to substitute the factor
$D_N(\eta)$ (see (\ref{1.4})) with its averaged value defined in
(\ref{1.9}). 
The sum on the r.h.s. of (\ref{1.9}) is carried out straightforwardly
resulting in the following expression for 
$\left\langle D_N(\eta)\right\rangle$:
 \begin{eqnarray}
\fl
\left\langle D_N(\eta)\right\rangle 
=
\exp\left(-{N \over N_d}\right)\,
\left({\sin N\pi \eta \over \sin \pi \eta}  \right)^2
+
{2 N_d^2 \left(1+(2N_d)^{-1}\right) \over 1 + 4N_d^2 \sin^2 \pi \eta}
\,
\left(1-\exp\left(-{N \over N_d}\right)\right)
\label{2.17}\\
\fl
\quad
-
\exp\left(-{N \over N_d}\right)\,
{N_d \over 1 + 4N_d^2 \sin^2 \pi \eta}
\,
{\sin N\pi \eta \over \sin \pi \eta}
\left[
2\cos(N+1)\pi \eta + 
{\sin (N+1) \pi \eta \over N_d \, \sin \pi \eta}\, \cos \pi \eta 
\right]
\nonumber
\end{eqnarray}
 Although more cumbersome, if compared with simple formula
(\ref{1.4}), the structure of  r.h.s. of (\ref{2.17}) can be readily
analyzed.

As well as the factor $D_N(\eta)$, the function $\left\langle
D_N(\eta)\right\rangle$ has sharp maxima at the points
$\eta=k=1,2,3\dots$ which correspond to the frequencies $\omega_k$
defined as in (\ref{1.4a}).  For given $N$ the maximum value $D_N(k)$,
which is the same for all $k$, is equal to
\begin{equation}
\fl
\left\langle D_N(k) \right\rangle 
=
N_d^2 \left( 2 + {1 \over N_d}\right)
\left[1 - \left(1+{N \over N_d}\right)
\exp\left(-{N\over N_d}\right) \right]
=
\cases
{
N^2,& for $N/N_d \ll 1$\\
2 N_d^2, & for $N/N_d \gg 1$\\
}
\label{4.3}
\end{equation}
and monotonously increases with the number of the undulator periods.

The limit $N/N_d \ll 1$ corresponds to a thin crystal, $L\ll L_d$.  In
this case the number of the channeling particles does not change
noticeably on the scale $L$, so that the AIR intensity is proportional
to $N^2$ as in an ideal undulator.  For $L \sim L_d$ the probability
of the particle to channel through the whole crystal length decreases,
and the value of $\left\langle D_N(k) \right\rangle$ starts to deviate
from the $N^2$-law.  In the limit of very thick crystal, $L \gg L_d$,
$\left\langle D_N(k) \right\rangle$ reaches its maximum value $2
N_d^2+N_d \approx 2 N_d^2$.

The r.h.s. of (\ref{2.17}) reduces to simple expressions in two
limiting cases of thin and infinitely thick crystals:
\begin{equation}
\left\langle D_N(\eta) \right\rangle 
=
\cases
{
\left({\sin N\pi \eta \over \sin \pi \eta}  \right)^2,
& for $N\ll N_d$\\
{2 N_d^2 \over 1 + 4N_d^2 \sin^2 \pi \eta},
& for $N\gg N_d$\\
}
\label{2.17a}
\end{equation}
Both limiting expressions have clear physical meaning.  For $N\ll N_d$
($L\ll N_d$) the dechanneling is of no importance, and, therefore, the
structure of the characteristic line of the AIR is that of an ideal
undulator, eq. (\ref{1.4}).  With $N$ ($L$) increasing the profile of
the characteristic line changes and, finally, becomes of the
Lorenz-type in the limit $N\gg N_d$ and for $|\eta-k|\ll 1$.  The
qualitative explanation of this result can be given if one interprets
the relation (\ref{1.5}) in terms of quantum mechanics.  Then, the
r.h.s. is proportional to the squared modulus of the projectile
wave-function which corresponds to the bound (channeled) state
characterized by a complex energy.  Indeed, the wave function of the
channeled particle can be presented in the form (e.g. \cite{Bazylev})
$\Psi({\bf r},t)=\exp(-\i\varepsilon t)\psi({\bf
r}_{\parallel})\psi({\bf r}_{\perp},t)$, where $\psi({\bf
r}_{\parallel})$ corresponds to the unbound longitudinal motion of the
particle, and $\psi({\bf r}_{\perp},t)$ describes the transverse
motion in the channel.  Provided the multiple scattering is excluded
the function $\psi({\bf r}_{\perp},t)=\psi({\bf
r}_{\perp})\exp(-\i\varepsilon_{\perp}t)$ corresponds to a stationary
bound state of the transverse motion ($\varepsilon_{\perp}$ is the
energy associated with the transverse degree of freedom).  In this
case, the normalization condition $\int \d {\bf r}_{\perp}|\psi({\bf
r,t}_{\perp})|^2 =1$ means, that the probability to find the particle
inside the channel does not depend on time $t$, which, in turn, is
related to the penetration distance through $t=z/c$.  Random
scattering by the electrons and nuclei can be incorporated in this
picture by adding the imaginary term to the energy of the transverse
motion, $\varepsilon_{\perp}\rightarrow \varepsilon_{\perp}+\i
(\Gamma/2) t$, where $\Gamma$ is the total width associated with the
transitions to the unbound (dechanneled) continuum due to the multiple
scattering.  Hence, the normalization condition acquires the form
$\int\d{\bf r}_{\perp}|\psi({\bf r,t}_{\perp})|^2=\exp(-\Gamma
t)=\exp(-\Gamma z/c)$.  Comparing this expression with (\ref{1.5}) one
finds $\Gamma = c/L_d \propto 1/N_d$ which is exactly the quantity
which defines the shape of the resonant line of $\left\langle
D_N(\eta) \right\rangle$ in the case $L\gg L_d$ (the second relation
in (\ref{2.17a})).

\begin{figure}
\hspace{2.5cm}\epsfig{file=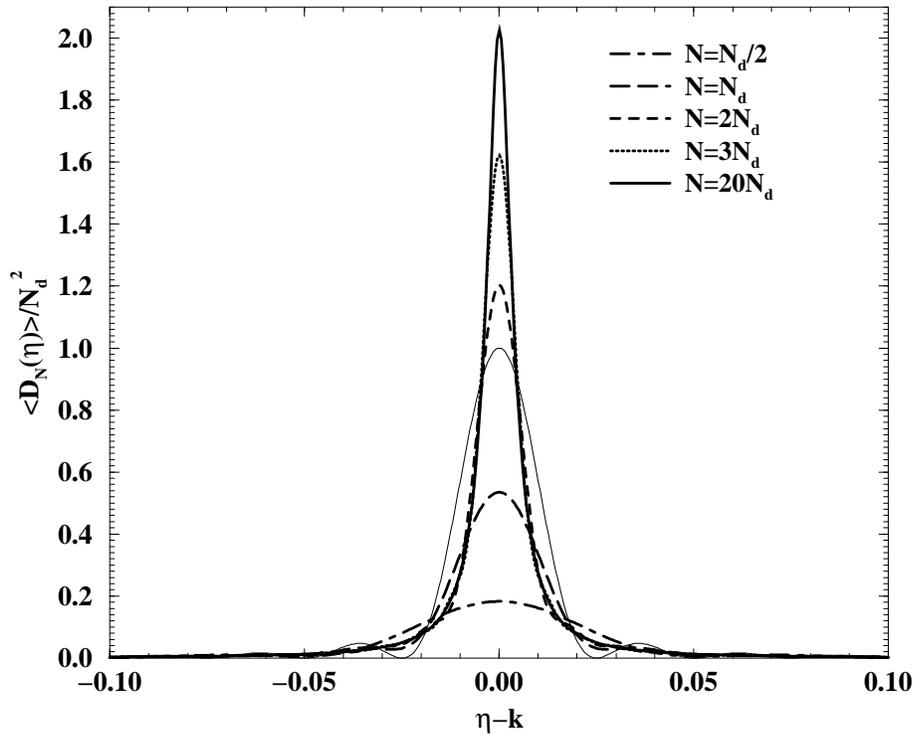,width=12cm, angle=0}
\caption{The dependences  $\left\langle D_N(\eta) \right\rangle/N_d^2$ 
(eq. (\protect\ref{2.17})) for various $N$ as indicated in vicinity of
the point $\eta=k$, ($k=1,2,\dots$). The parameter $N_d$ is set to 40.
The thin solid line corresponds to $D_N(\eta)/N^2$ (see
(\protect\ref{1.4})) calculated for $N=N_d$.}
\label{fig.DN}
\end{figure}

The change in the shape of the characteristic line is illustrated in
figure \ref{fig.DN} where the function $\left\langle D_N(\eta)
\right\rangle/N_d^2$ is plotted for several values of $N$ (as
indicated) in the vicinity of the point $\eta \approx k$.  For the
sake of comparison we plotted also the dependence $D_N(\eta)/N^2$ for
$N=N_d$ (the thin solid line) which characterises the line shape in an
ideal undulator without the decrease in the volume density of the
channeled particles.
 
Hence, the expression (\ref{1.8}) combined with (\ref{2.17}) describes
the spectral-angular of the total radiation in the range $\omega \sim
\omega_k \ll \omega_{ch}$ formed during a passage of an
ultra-relativistic positron along periodically bent crystallographic
plane.  All the complexity of the trajectory related to the channeling
oscillations are incorporated in the undulator parameter $p^2$ (see
(\ref{p2_mod})).  The effect of the dechanneling is included in the
factor $\left\langle D_N(\eta) \right\rangle$, eq. (\ref{2.17}), which
depends on two parameters, $N$ and $L_d$.  The number of the undulator
periods $N$ can be varied by changing the length of the crystal $L$
or/and by changing the period $\lambda$ of the shape function $S(z)$
(see figure \ref{fig.1st})).

In the case of a straight channel the dechanneling length $L_d$
depends on the type of a crystal and on the energy of projectile
\cite{Gemmell,KumakKomar}.  For a crystal bent with constant curvature
$1/R$ the quantity $L_d$ becomes also dependent on $R$.  The methods
for estimating $L_d$ for heavy projectiles channeling in crystals bent
with $1/R=const$ are discussed in \cite{Biryukov}.  To our knowledge,
no detailed study has been carried out of the dependence of the
dechanneling length on the parameters of the shape function $S(z)$ in
the case of a light projectile (a positron) channeling through
periodically bent crystal.  In the next two sections we present the
approach and the results of numerical calculation of $L_d$ in this
case.
  
\section{Simulation of the dechanneling process in periodically 
bent crystals}
\label{Dechanneling}

As it was demonstrated above the intensity of the radiation and the
profile of the line of the AIR are defined, to a great extent, by the
parameter $L_d$.

This section is devoted to the description of the model which we used
to calculate the dechanneling length $L_d$ for an ultra-relativistic
positron channeling through periodically bent crystal.  Our approach
is based on the simulation of the trajectories and the dechanneling
process of an ultra-relativistic positron.  This is done by solving
the three-dimensional equations of motion which account for (i) the
interplanar potential, (ii) the centrifugal potential due to the
crystal bending, (iii) the radiative damping force, (iv) the
stochastic force due to the random scattering of projectile by lattice
electrons and nuclei.

We analyze the dependence of $L_d$ on the energy of the projectile,
the type of the crystal and crystallographic plane, the parameters of
the shape function $S(z)$. The latter include two factors: the
parameter $C$ and the ratio $a/d$ (see
eqs. (\ref{cond1}-\ref{cond2})).

The results of our calculations of the quantity $L_d$, presented in
section \ref{Numerical1}, are used further in section \ref{specrtaAIR}
to calculate the spectral-angular and the spectral distributions of
the AIR.

\subsection{Equations of motion in periodically bent crystal with
account for the radiation damping}
\label{EM}

Below we outline the derivation of the equations of motion for an
ultra-relativistic positron channeled in a periodically bent crystal.
Contrary to the case of a heavy projectile, a light projectile (a
positron, an electron) with $\varepsilon \geq 1600\, m/Z$ looses its
energy, when passing through matter, mainly due to the radiative
losses (e.g. \cite{Akhiezer}).  In this inequality $Z$ is the atomic
number of the crystal atoms.  Hence, the radiative losses exceed the
losses due to the ionizing collisions starting with $\approx 60$ MeV
in Si, and $\approx 11$ MeV in W.  Therefore, accurate treatment of
the equations of motion of the channeled ultra-relativistic positron
must account for the effect of the radiation damping
\cite{Bonch,Huang,Baier97}.
In our scheme by introducing the classical 
radiative damping force  \cite{Land2}.

The equations of motion for a relativistic particle of mass $m$ and
charge $e$ moving in an external static electric field ${\bf E}$ read
as follows
\cite{Land2}:
\begin{eqnarray}
\left\{ 
\begin{array}{rcl} 
m\, {\d \over \d t}(\gamma {\bf v})
&=&
e {\bf E} + {\bf f}
\\
mc^2\, \dg 
&=&
e {\bf E}\cdot{\bf v} + {\bf f}\cdot{\bf v}
\end{array} 
\right. 
\label{my.16a}
\end{eqnarray}
Here
\begin{equation}
{\bf f} 
= {2 e^3 \over 3 m c^3}\, 
\left[
\gamma\, ({\bf v}\bbnabla)\, {\bf E}
+
{e \over m c^2}\,
({\bf v}{\bf E})\, {\bf E}
-
{e \over m c^2}\,
\gamma^2\,
\left(
{\bf E}^2-
{({\bf v}{\bf E})^2 \over c^2}
\right)
\,{\bf v} 
\right]
\label{my.10}
\end{equation}
is the radiative damping force due to the presence of the electric
field ${\bf E}$.  The latter is related to the interplanar potential
$U({\bf r})$ through $e {\bf E} = - \partial U/\partial {\bf r}$,
where ${\bf r}=y\,{\bf e}_y + z\, {\bf e}_z + x\, {\bf e}_x$ is the
radius vector.

\begin{figure}
\hspace{2.5cm}\epsfig{file=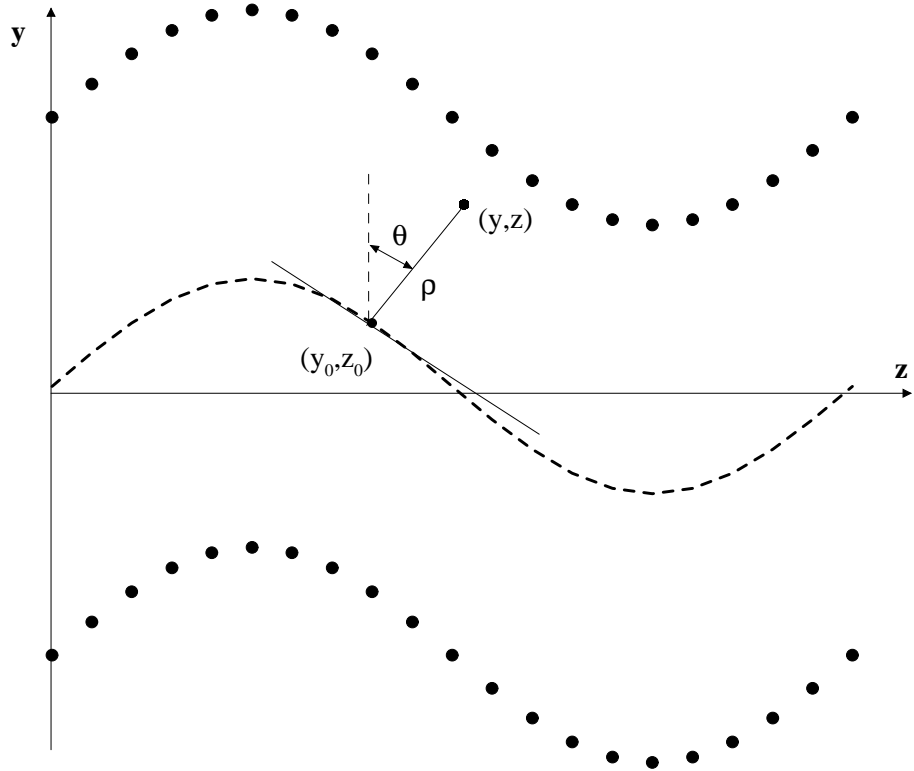,width=12cm, angle=0}
\caption{The coordinate $\rho$ in a bent channel: $\rho$ is the
distance from some point $(y,z)$ in the channel to its centerline. To
obtain $\rho$ it is necessary to find the point $(y_o,z_o)$ on the
centerline in which the tangent to the centerline is perpendicular to
$\vec{\rho}$.  If omitting the terms $S^{\prime\, 2}(z)$ and higher
the $\rho$ coordinate equals $\rho = y- S(z)$.}
\label{fig.bchan2}
\end{figure}

Let the crystal centerline be bent as it is presented in figure
\ref{fig.1st}.  We impose the following condition on the derivative of
the periodic profile function $S(z)$:
\begin{equation}
 S^{\prime\, 2}(z)\sim (a/\lambda)^{2} \equiv \xi \ll 1
\label{condS}    
\end{equation}
Then, the length $L$ of the bent crystal, the interplanar spacing $d$,
and the local curvature $1/R(z)$ of the centerline satisfy the
relations
\begin{equation}
\fl
L = L_0(1 + O(\xi^2)) \approx L_0,
\quad
d_0 (1 + O(\xi^2)) \approx d_0,
\quad
R^{-1} = 
|S^{\prime\prime}(z)| (1 + O(\xi^2)) \approx
|S^{\prime\prime}(z)| 
\label{LdR}    
\end{equation}
Here the subscript $0$ refers to the straight channel.  When combined,
the relations (\ref{LdR}) lead to the conclusion that the parameters
of the bent channel, which are the channel width and the distance
between the atoms of any crystallographic plane, are equal to the
their values in the case of straight crystal provided one neglects the
terms of the order $\xi^2$ and higher.  This, in turn, allows to
formulate the following two conditions regarding to the interplanar
potential in periodically bent crystal:
\begin{itemize}
\item[1.]  within any single bent channel the
potential $U$ depends only on the variable $\rho\approx y-S(z)$ which
is the distance from the $(x,y,z)$ point to the channel centerline
(see figure \ref{fig.bchan2});
\item[2.] the dependence of $U$ on $\rho$ in the 
 bent channel is identical to that in the linear one. 
Therefore,
\begin{equation}
U({\bf r}) = U(\rho),
\qquad
\rho = y -S(z)
\label{bent_pot}    
\end{equation}
\end{itemize}

Eq. (\ref{bent_pot}) leads to the following definition of the
interplanar static electric field ${\bf E}$:
\begin{eqnarray}
e {\bf E} = 
-\dU\,
\left({\bf e}_y- \dS\,  {\bf e}_z\right)
\label{2.2a}
\end{eqnarray}

Using (\ref{2.2a}) in (\ref{my.10}) and in (\ref{my.16a}) and carrying
out some algebra one ends up with the following system of coupled
equations
\begin{eqnarray}
\ddr 
&=&
-
\left(
{\dU \over m\gamma}
+
c^2\ddS 
+
{1 \over m}\,
{2 r_0 \over 3 c}\,
\dr\,\ddU 
\right)
\label{4.8a}\\
\ddx
 & = & 
{\dU \, \dr\, \dx \over m\gamma c^2}
\label{4.8x}\\
\dg
 & = & 
-{1 \over mc^2}\,
\left(
\dU\, \dr 
+
{2 r_0 \over 3 c}\,
{\gamma^2\, {\dU}^2 \over m}\, 
\right)
\label{4.8b}
\end{eqnarray}
Here $r_0=e^2/mc^2=2.818\times 10^{-13}$ cm is the classical electron
radius.

Since $1-\gamma^{-2}=(\dx^2+\dy^2+\dz^2)/c^2$ then dependence $z(t)$
is uniquely defined provided the functions $\rho(t)$, $x(t)$ and
$\gamma(t)$ are known.

On the r.h.s. of (\ref{4.8a}-\ref{4.8b}) the terms of the order
$\gamma^{-2}$ are omitted.  We also dropped out the terms of the order
$U/m\gamma c^2$. This is justified by the fact that for most of
crystals the depth of the interplanar potential well $U_0$ lies within
$10\dots 100$ eV (see e.g. \cite{Baier}).  Therefore, for $1\dots 10$
GeV positrons the parameter $U_0/m\gamma c^2 \sim 10^{-9}\dots 10^{-7}
\leq \gamma^{-2}$.

The profile of the periodically bent channel enters the equations
(\ref{4.8a}-\ref{4.8b}) via the function $\ddS(z)$.  Recalling
(\ref{LdR}), the second term on the r.h.s. of (\ref{4.8a}) takes the
form $-c^2/R(z)$ and, thus, represents the acceleration due to the
centrifugal force.

\subsection{Equations of motion with account for 
the random collisions with target electrons and nuclei}
\label{EM1}

Random scattering of a projectile by target electrons and nuclei leads
to deviation of the trajectory from that which is obtained by solving
equations of motion (\ref{4.8a}-\ref{4.8b}).  There are two different
approaches aimed to account for these deviations.  The first approach
is based on the diffusion theory applied to describe the multiple
scattering (e.g. \cite{Kumakhov1986,KumakKomar,Taratin1980,Andersen}.
The second approach is in direct computer simulation of the scattering
process \cite{Biryukov95,Biryukov}.  In the present paper we adopt the
scheme similar to the one described by Biryukov
\cite{Biryukov95,Biryukov}.

Scattering from target electrons results in two changes in the
projectile motion.  The first one is the gradual decrease in the
projectile energy due to the ionization losses.  The second effect of
the collisions is that they lead to a chaotic (random) change in the
direction of the projectile motion.

The scattering from target nuclei results in a chaotic (random) change
in the direction of the projectile motion.

\subsubsection{Ionization losses for a channeling positron}

Although for an ultra-relativistic positron ($\gamma\gg 1,\ v\approx
c$) passing through media the ionization losses are much smaller as
compared with the radiative ones (see e.g. \cite{Land4,Komarov}) they
are incorporated into the scheme which is described here.  The mean
energy loss per path $\d s \approx c \d t$ due to the electronic
scattering can be written as a function of the distance $\rho$ from
the midplane
\cite{Biryukov,Land4,Komarov,Rossi}
\begin{equation}
\fl
\left(- {1 \over c} {\d \varepsilon \over \d t}\right)_{ion}
\!\!\!\! =
K m c^2 \langle n_{el}\rangle
\,
\left[
{\rm ln}{\gamma \sqrt{2m c^2 \, T_{{\rm max}}}\over I} - {23 \over 24}
-{\delta \over 2} + C(\rho) 
+
{n_{el}(\rho)\over \langle n_{el}\rangle}
\left(
{\rm ln}{T_{{\rm max}} \over I} - {1 \over 2}
\right)
\right]
\label{6.5}
\end{equation}
The notations used here are as follows:
\begin{itemize}

\item
$K=2\pi r_0^2 \approx 5\times 10^{-25}$ cm$^2$. 

\item
The quantity $\langle n_{el}\rangle $ is the mean concentration of
electrons in the corresponding amorphous media.  It is related to the
mass density $\rho_m$ of the material, the atomic weight of the
crystal $A$ and the atomic number $Z$ of its atoms through $\langle
n_{el}\rangle = N_A\, Z\rho_m/A$ (with $N_A = 6.022\times 10^{23}$
mol$^{-1}$ being the Avogadro number).

\item
$n_{el}(\rho)$ is the local concentration of the electrons inside the
channel.  To calculate $n_{el}(\rho)$ we use the Moli\`ere
approximation.

\item
The quantity $\delta$ is a so-called density effect correction
\cite{Komarov,Sternheimer}. 
For an ultra-relativistic positron one can use $\delta \approx 2{\rm
ln}\gamma$ \cite{Land4}.

\item
The quantity $T_{{\rm max}}$ is the maximum magnitude of the energy
transfer from a projectile to a target electron.  For an
ultra-relativistic positron $T_{{\rm max}} \approx \gamma m c^2$.

\item
The term $C(\rho)$ is the correction due to the periodicity of the
crystalline structure. The exact form of this correction was found in
\cite{Golovchenko}. 

\end{itemize}

Substituting $\varepsilon = \gamma mc^2$ and $T_{{\rm max}}=\gamma m
c^2$ into (\ref{6.5}) we write
\begin{equation}
\fl
\left({\d \gamma \over \d t}\right)_{ion}
=
- K c\, \langle n_{el}\rangle
\,
\left[
{\rm ln}{m c^2 \sqrt{2\gamma}\over I} - {23 \over 24}
+ C(\rho) 
+
{n_{el}(\rho)\over \langle n_{el}\rangle}
\left(
{\rm ln}{\gamma m c^2 \over I} - {1 \over 2}
\right)
\right]
\label{6.5a}
\end{equation}
To account for the influence of the ionization losses on the
channeling motion one can add (\ref{6.5a}) to the r.h.s. of
(\ref{4.8b}).  Then, the resulting equation for $\gamma$ reads
\begin{eqnarray}
\dg
=  
-{1 \over mc^2}\,
\left(
\dU\, \dr 
+
{2 r_0 \over 3 c}\,
{\gamma^2\, {\dU}^2 \over m}\, 
\right)
+
\left({\d \gamma \over \d t}\right)_{ion}
\label{4.12b.2}
\end{eqnarray}
 
\subsubsection{Random change in the direction of motion due to ionizing
collisions}
\label{YesElCollision1}

To account for the random change in the direction of motion due to
single collisions with target electrons we followed the procedure
proposed by Biryukov \cite{Biryukov95} for heavy projectiles but
modifying his formalism for the case of a light projectile.  For an
ultra-relativistic positron travelling through a crystal the
differential probability (per path $\d s=c\d t$) of the relative
energy transfer $\mu = (\varepsilon-\varepsilon^{\prime})/\varepsilon$
due to the the ionizing collisions with the quasi-free electrons is
defined by the following expression
\cite{Rossi,DataGroup}:
\begin{equation}
{\d^2 P \over \d \mu\, \d s }
=
{1 \over \gamma}\, {K\, n_{el}(\rho) \over \mu^2}\, 
\label{6.1}
\end{equation}

The relative energy transfer defines a round scattering angle
$\theta$, measured with respect to the instant velocity of the
projectile, through (see e.g. \cite{Land2})
\numparts
\begin{eqnarray}
\cos\theta
=
\sqrt{
{1 - \mu/\mu_{{\rm max}}\over 1 - \alpha\,\mu/\mu_{{\rm max}}}
}
\label{6.3a}\\
\mu \in [\mu_{{\rm min}},\mu_{{\rm max}}],
\qquad
\mu_{{\rm max}} = 1 - {1\over \gamma},
\qquad
\mu_{{\rm min}} = I /\varepsilon
\label{6.3b}\\
\alpha= {\gamma - 1 \over \gamma + 1}\approx 1- {2\over \gamma} 
+ O\left(\gamma^{-3}\right)
\label{6.3c}
\end{eqnarray}
\endnumparts
Here $I\approx 16\, Z^{0.9}$ eV characterises the average ionization
potential of the crystal atoms.  The maximum relative energy transfer
$\mu=\mu_{{\rm max}}$ results in $\theta=\pi/2$, whereas
$\mu=\mu_{{\rm min}}\ll 1$ leads to nearly forward scattering.

Following Biryukov \cite{Biryukov95} we used the distribution
(\ref{6.1}) to generate (randomly) the magnitude of $\theta$ when
integrating the system of equations (\ref{4.8a}-\ref{4.8x}) and
(\ref{4.12b.2}).  The magnitude of the second (the azimuthal)
scattering angle $\phi$ is not restricted by any kinematic relations,
and is obtained by random shooting (with a uniform distribution) into
the interval $[0,2\pi]$.

The algorithm of random generation of the $\theta$-values is as
follows.  The two-fold probability $\d P/ \d \mu\, \d s$ satisfies the
normalization condition
\begin{equation}
\int_0^{L_{{\rm ion}}}\, \int_{\mu_{{\rm min}}}^{\mu_{{\rm max}}}\, 
{\d^2 P \over \d \mu\, \d s }
\, 
\d \mu\, \d s 
= 1
\label{6.1.a}
\end{equation}
where $L_{{\rm ion}}$ is the interval inside which the probability of
the projectile to undergo the ionizing collision accompanied by
arbitrary energy transfer is equal to 1.  The parameter $L_{{\rm
ion}}$ (defined from (\ref{6.1.a})) is expressed through $\mu_{{\rm
min}}$, $\mu_{{\rm max}}$ and the local electron density
$n_{el}(\rho)$ as follows
\begin{equation}
L_{{\rm ion}}^{-1}=
{K\, n_{el}(\rho) \over \gamma}\, {1 \over \mu_{{\rm min}}},
\label{6.1.b}
\end{equation}

Using these notations eq. (\ref{6.1}) can be  written as
\begin{equation}
\d P 
=
{\Delta s \over L_{{\rm ion}}}\,
W(\mu)\,d \mu 
\qquad
W(\mu)
=
{\mu_0 \over \mu^2} 
\label{6.1.c}
\end{equation}
Here the factor $\Delta s/L_{{\rm ion}}$ defines the probability of
the collision (any) to happen on the scale $\Delta s$, whereas the
factor $W(\mu)\, \d \mu$ represents the normalized probability of the
energy transfer between $\mu$ and $\mu + \d \mu$.  To generate random
deviate with the probability distribution $W(\mu)$ we used the
algorithm described in \cite{NumRec}.

The scheme outlined above implies that the ionizing collisions are
treated as events.  Hence, for each step $\Delta s = c\,\Delta t $ of
integration of the the system (\ref{4.8a}-\ref{4.8x}, \ref{4.12b.2})
we first simulate the probability of the event to happen by generating
a uniform random deviate $x\in [0,1]$ and comparing it with $\Delta
s/L_{{\rm ion}}$.  If $x \leq \Delta s/L_{{\rm ion}}$ then the
scattering angles $\theta$ and $\phi$ are calculated and used to
modify the direction of motion of the projectile but leaving the
magnitude of the projectile velocity unchanged.

\subsection{Random change in the direction of motion due to scattering 
from nuclei}
\label{YesNuclCollisions}

The change in the projectile direction of motion due to the collisions
with crystal nuclei was taken into account at every step when
integrating the system (\ref{4.8a}-\ref{4.8x}), (\ref{4.12b.2}).  The
square of small deflection angle $\Delta \theta^2$ per path $\Delta
s$, was computed from a Gaussian distribution
\begin{equation}
{\d P \over \d \kappa}
=
{1 \over \sqrt{2\pi \overline{\kappa^2}}}\, 
\exp\left(- {\kappa^2 \over 2\,\overline{\kappa^2} }\right),
\qquad
\kappa^2 = {\Delta \theta^2 \over \Delta s }
\label{nucl.1}
\end{equation}
Here $\overline{\kappa^2} \equiv \overline{\theta^2}/\Delta s $ is the
mean-square deflection angle due to the scattering from nuclei in
crystals.  In the approximation suggested by Kitagawa and Ohtsuki
\cite{Kitagawa}
\begin{equation}
\overline{\kappa^2} = 
{ n_{n}(\rho)\over \langle n_{n}\rangle}\,
\langle \kappa^2 \rangle 
\label{nucl.2}
\end{equation}
where $\langle n_{n}\rangle$ is the mean concentration of nuclei in an
amorphous media, $n_{n}(\rho)$ is the local concentration of the
nuclei inside the channel.  With the thermal vibrations taken into
account the quantity $ n_{n}(\rho)$, corresponding to the distribution
of the nuclei of two neighbouring planes versus the distance from the
midplane, is given by
\begin{equation}
{ n_{n}(\rho)\over \langle n_{n}\rangle}=
{ d \over  \sqrt{2\pi u_T^2}}\,
\left[ 
{\rm exp}\left(-{(d/2+\rho)^2 \over 2\, u_T^2}\right)
+
{\rm exp}\left(-{(d/2-\rho)^2 \over 2\, u_T^2}\right)
\right]
\label{nucl.3}
\end{equation}
with $T$ standing for the crystal temperature and $u_T$ denoting the
thermal vibration root-mean-square amplitude.

The quantity $\langle \kappa^2 \rangle$ on the r.h.s. of
(\ref{nucl.2}) is the mean-square of $\Delta\theta^2/\Delta s $ in an
amorphous media.  For an ultra-relativistic projectile it is given by
\cite{KumakKomar}
\begin{equation}
\langle \kappa^2 \rangle
=
{\varepsilon_s^2 \over \varepsilon^2}\,
{1 \over L_{{\rm rad}}}
\label{nucl.4}
\end{equation}
where $\varepsilon_s=21$ MeV, and $L_{rad}$ is the radiation
length \cite{Land4}:
\begin{equation}
L_{{\rm rad}}^{-1}
\approx
4\alpha\, r_0^2 \, Z^2\, \langle n_{n}\rangle\, 
{\rm ln} \left(183\,Z^{-1/3}\right)   
\label{nucl.5}
\end{equation}

\section{Numerical results of the dechanneling process
simulations}
\label{Numerical1}

We used the scheme outlined in sections 
\ref{EM} and \ref{EM1} for computer modelling of the 
dechanneling process of ultra-relativistic positrons in periodically
bent crystals.  The goal of these calculations was to obtain realistic
dependences of the number of channeled particles $n(z)$ versus the
penetration distance $z$.  The interplanar potential was considered
within the the Moli\`ere approximation at the temperature $T=150$ K.
 
The calculations were performed for 5 GeV positrons channelling along
the $(110)$ crystallographic planes in Si, Ge, and W crystals.  The
shape function $S(z)$ was chosen in the form (\ref{sinSz}).  In this
case the minimum curvature radius of the centerline bending equals
$R_{\rm min}= 1/(a k^2)=\lambda^2/4\pi^2 a$, and, consequently, the
parameter $C$ defined in (\ref{cond1}) reads as $C = \varepsilon a
k^2/ \dU_{{\rm max}}$.  Thus, for given crystal and crystallographic
plane, and for given $\varepsilon$ the parameter $C$ and the ratio
$a/d$ uniquely define the period $\lambda$.

In our calculations $C$ was varied from 0 to 0.5, and the considered
ratios $a/d$ are: 0 (the case of a straight channel), 5,10, 15 and 20.

For each pair of the parameters $C$ and $a/d$ we simulated 2000
trajectories by solving the coupled system (\ref{4.8a}-\ref{4.8x}) and
(\ref{4.12b.2}), which was integrated with the account for the random
scattering from target electrons and nuclei as explained in section
\ref{EM1}.  The initial conditions were chosen as follows. The
coordinates $x$ and $z$ were set to zero at the entrance, the
transverse coordinate $\rho$ was obtained by random shooting (with a
uniform distribution) into the interval $[-d/2, d/2]$.  The vector of
initial velocity ${\bf v}_0$ was aligned with the tangent to the
centerline at the entrance but with allowance for the spread in the
incident angles $\Theta \in [-\Theta_{\rm max}, \Theta_{\rm max}]$.
Hence, the initial values of the velocities $\dx/c$ and $\dr/c$ were
obtained by random shooting (with a uniform distribution) into the
interval $[-\Theta_{\rm max}, \Theta_{\rm max}]$.  For each crystal
the parameter $\Theta_{\rm max}$ was chosen as $10^{-4}\Theta_{\rm
L}$.  The values for the Lindhard angle $\Theta_{\rm L}$ for 5 GeV
positron channeling along (110) planes in Si, Ge and W are equal to
$9.57\times10^{-5}\,, 1.28\times10^{-4}\, ,2.35\times10^{-4}$, rad,
respectively.  Thus, the considered spread in the incident angles is
about $10^{-8}$ rad which is achievable in modern high-energy $e^{+}\,
e^{-}$ colliders (see e.g. \cite{DataGroup}).

\begin{figure}
\hspace{2.5cm}\epsfig{file=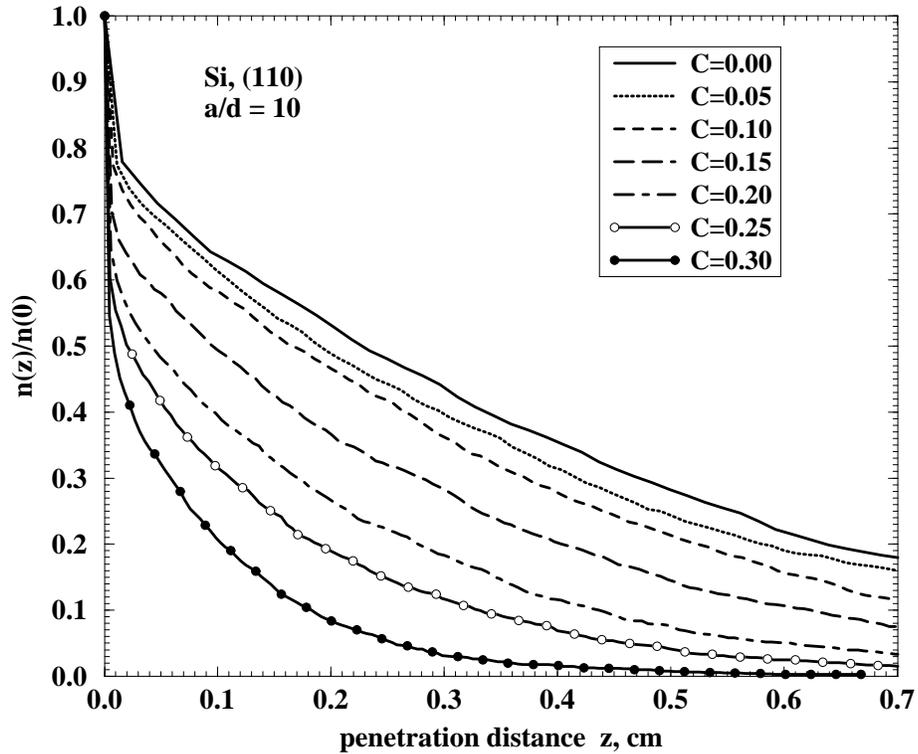,width=12cm, angle=0}
\caption{The calculated dependences $n(z)/n(0)$ versus penetration 
distance $z$ for $5$ GeV positrons channeling along the $(110)$ in Si
crystal for various values of the parameter $C$ (see
(\protect\ref{cond1})) as indicated.  The data correspond to the shape
function $S(z)=a\sin(2\pi z/\lambda)$.  The $a/d$ ratio equals $10$.
For each indicated $C$ the corresponding values of $\lambda$, and the
calculated magnitudes of the dechanneling lengths $L_d^c$ and the
number of undulator periods $N_d^c=L_d^c/\lambda$ are presented in
Table \protect\ref{Table1}.  See also the commentaries in the text.}
\label{fig.Ld_si}
\end{figure}

\begin{figure}
\hspace{2.5cm}\epsfig{file=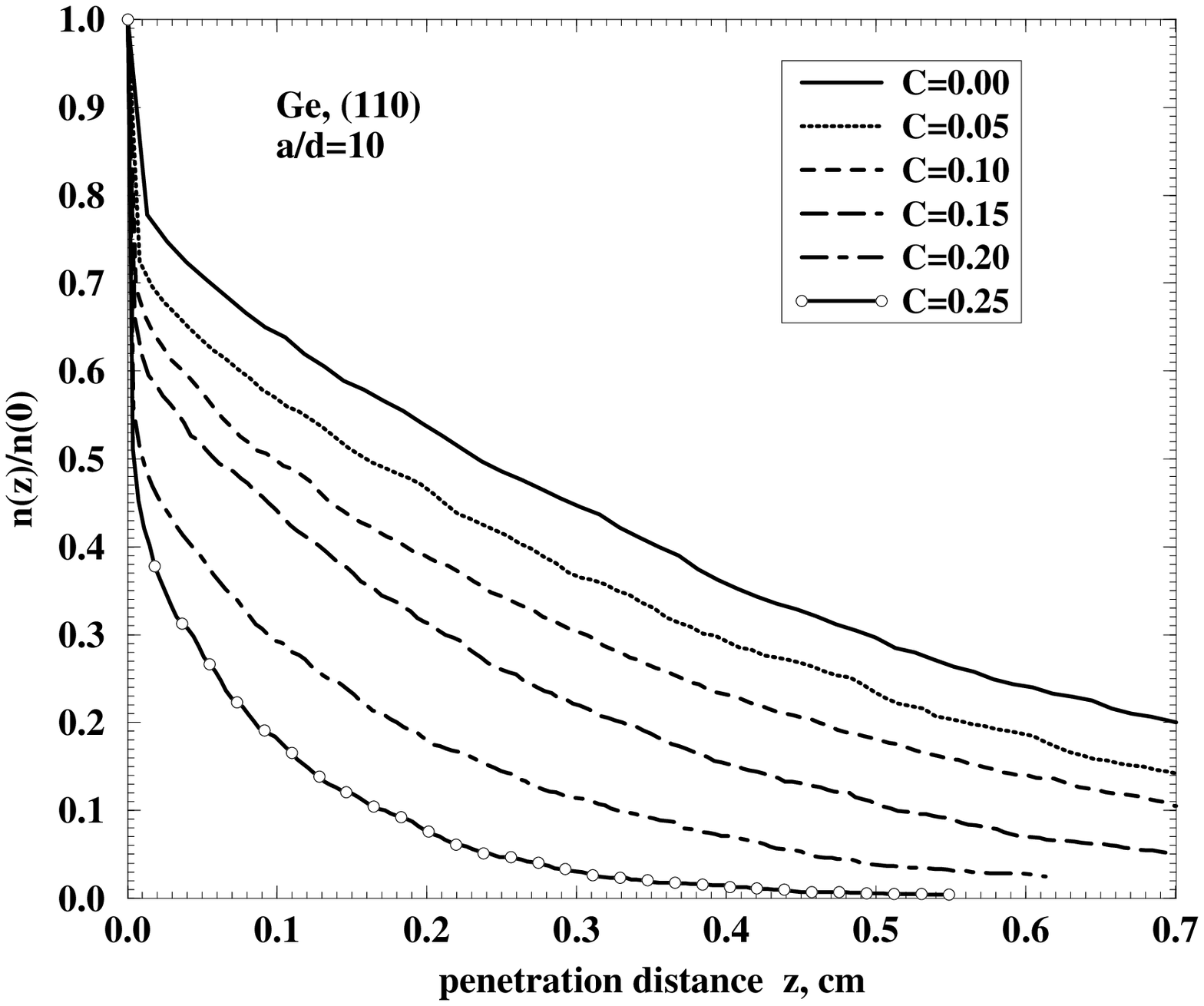,width=12cm, angle=0}
\caption{Same as in fig. \protect\ref{fig.Ld_si} but for 
Ge crystal.} 
\label{fig.Ld_ge}
\end{figure}

\begin{figure}
\hspace{2.5cm}\epsfig{file=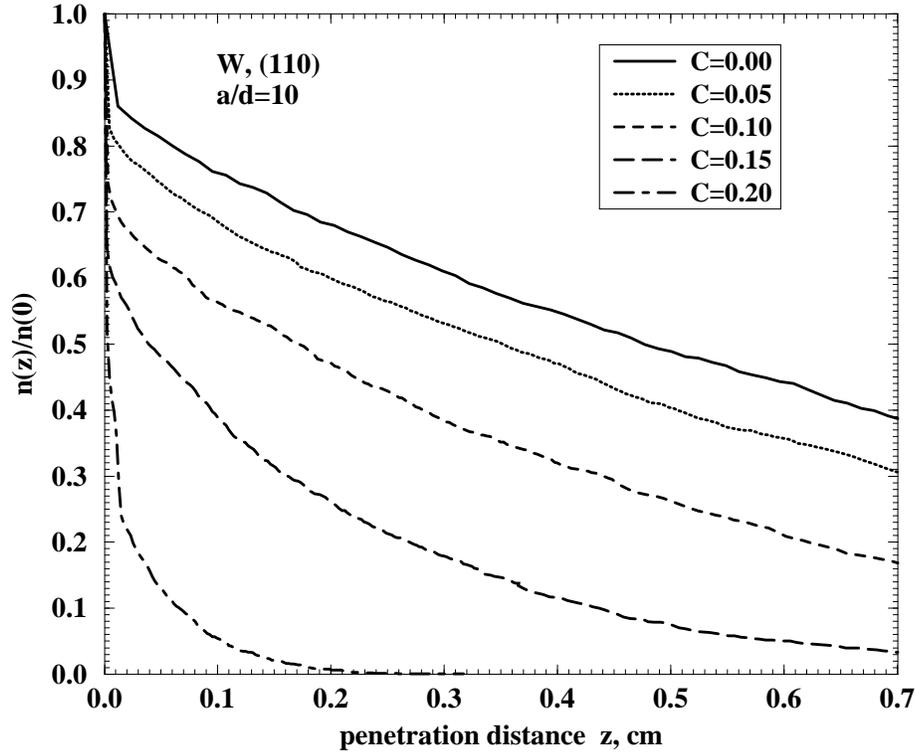,width=12cm, angle=0}
\caption{Same as in fig. \protect\ref{fig.Ld_si} but for 
W crystal.} 
\label{fig.Ld_w}
\end{figure}

The results of calculations are presented in figures
\ref{fig.Ld_si}-\ref{fig.Ld_w}, \ref{fig.Ld_si_all} and in table
\ref{Table1}.

The first three figures represent the dependences of $n(z)$
(normalized to the entrance value $n(0)=n_0$) versus the penetration
distance $z$ for Si, Ge and W crystals.  The ratio $a/d$ equals 10.
The curves in the figures refer to different values of $C$ as
indicated.  The value $C=0$ stands for the case of a straight channel.
The corresponding values of the spatial period $\lambda$ of the shape
function $S(z)$ can be calculated as $\lambda=2\pi \sqrt{(\varepsilon
d/\dU_{{\rm max}})(a/d)C^{-1}}$ and are presented in the table.

\begin{figure}
\hspace{2.5cm}\epsfig{file=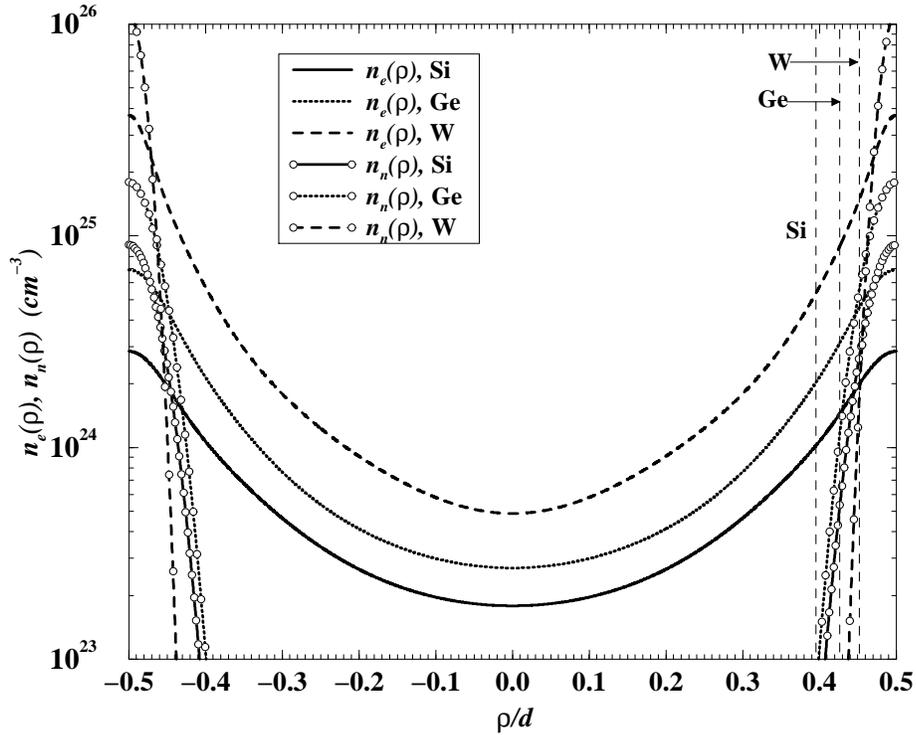,width=12cm, angle=0}
\caption{Volume densities of electrons and nuclei 
versus the distance $\rho$ from the midplane calculated within the
Moli\`ere approximation (at $T=150$ K) for (110) channels in Si, Ge
and W crystals.  The dashed lines mark the ratios $(d/2-a_{\rm
TF})/d$. The values of $d$ and $a_{\rm TF}$ for Si, Ge, W are equal
$1.92,\, 2.00,\, 2.45$ \AA\ and $0.194,\, 0.148,\, 0.112$ \AA,
respectively.}
\label{fig.NeNr}
\end{figure}

\begin{figure}
\hspace{2.5cm}\epsfig{file=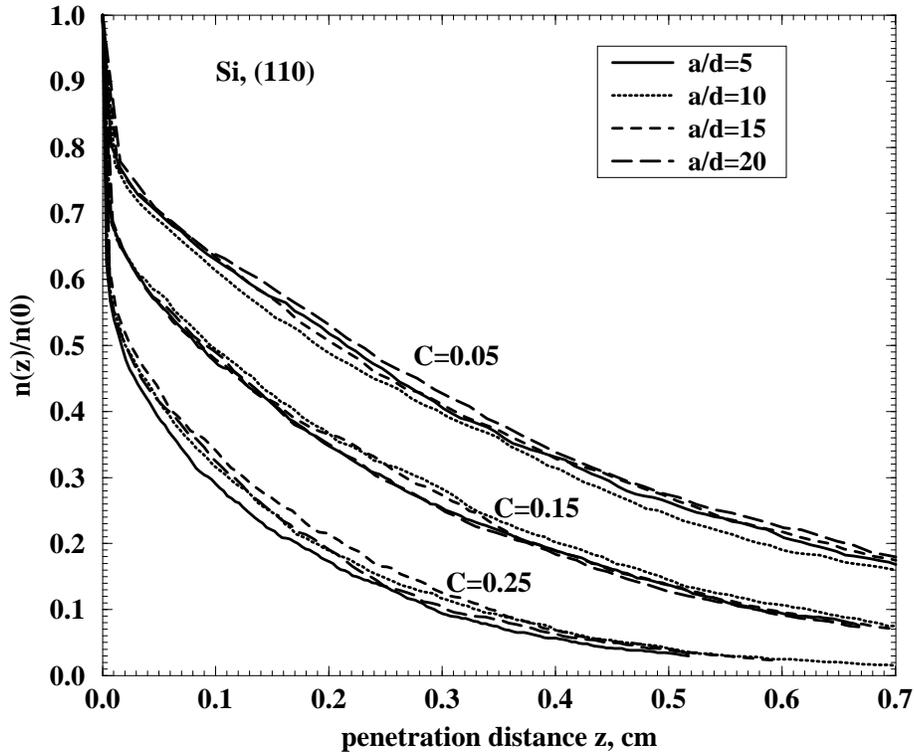,width=12cm, angle=0}
\caption{The calculated dependences $n(z)/n(0)$ versus penetration 
distance $z$ for $5$ GeV positrons channeling along the $(110)$ in Si
crystal for various values of the parameter $C$ and of the ratio $a/d$
as indicated.  The data correspond to the shape function
$S(z)=a\sin(2\pi z/\lambda)$.}
\label{fig.Ld_si_all}
\end{figure}

Let us discuss these results.  First to be mentioned is that in all
cases the dependences $n(z)/n(0)$ are monotonously decreasing
functions which are smooth for all $z$ except for the steep change in
the magnitude in the region close to $z=0$.  This change, $\Delta
n_0$, is the smallest for the straight channels and its magnitude
increases with $C$.  Such a behaviour is explained as follows.  The
probability for the projectile to undergo the large-angle scattering
from the crystal electrons or/and nuclei is proportional to the volume
densities $n_{el}(\rho)$ and $n_{n}(\rho)$ (see
(\ref{6.1.b}-\ref{nucl.3})).  The large-angle scattering, i.e. when
$\theta > \Theta_{\rm L}$, results in the increase of the kinetic
transverse energy, which becomes high enough for the projectile
positron to leave the potential well.  Hence, all those particles
which at the entrance move in the region of high nuclear and electron
densities leave the channeling mode almost immediately.  The volume
density of crystal electrons is comparatively small in the inner
region of the channel and sharply increases as $|\rho| > d/2 - a_{\rm
TF}$ (here $a_{\rm TF}=0.8853 Z^{-1/3} a_0$ is the Thomas-Fermi radius
of the crystal atoms, $a_0$ is the Bohr radius), the density of target
nuclei becomes noticeable for $|\rho| > d/2 - u_T$ ($u_T$ standing for
the thermal vibration root-mean-square amplitude).  This is
illustrated by figure \ref{fig.NeNr}.  Since in our simulations we
assumed that the particles are uniformly distributed in the coordinate
$\rho$ at the entrance point, then the fraction $\sim n(0)\,
(d-2\,a_{\rm TF})/d$ leaves the channel close to $z=0$.  This the
mechanism of dechanneling in the vicinity of the entrance point
\cite{Biryukov} is intrinsic for both straight ($C=0$) and bent
($C>0$) channels.  In addition to it, in the case of bent channel
there is a process which is called bending dechanneling
\cite{Biryukov}, and which is due to the change in the depth of the
potential well.  Indeed, inside the bent channel the transverse motion
of the particle is subject to the action of the effective potential
(see (\ref{4.8a}))
\begin{equation}
U_{eff}(\rho) 
= 
U(\rho) - {\varepsilon \ddS(z) }\, \rho
=
U(\rho) - {\varepsilon \over R(z)}\, \rho
\label{2.25}
\end{equation}
where and $R(z)$ is the (local) curvature radius of the channel.  For
a channel bent as described by (\ref{sinSz}) $R(z) =
\left[(\lambda/2\pi)^2/a\right]\,\sin(2\pi z/\lambda)$.  The particle
could be trapped into the channelling mode provided its total energy
associated with the transverse motion is less the minimal value,
$U_{eff}(\rho_{\rm min})$, of the two maxima points of the asymmetric
potential well described by (\ref{2.25}) \cite{losses,Biryukov}.  The
potential $U_{eff}(\rho)$ reaches the magnitude of $U_{eff}(\rho_{\rm
min})$ at some point $\rho_{\rm min}$ which satisfies the condition
$|\rho_{\rm min}|< d/2$ and the absolute value of $\rho_{\rm min}$
decreases with the growth of $a$.  Hence, the more the channel is
bent, the lower the allowed values of the channelling oscillations
amplitude are, and, consequently, the narrower is the interval of the
initial $\rho_0$ values for which the trajectory from the very
beginning will correspond to the channeling one.

Another feature to be mentioned in connection with the curves in
figures \ref{fig.Ld_si}-\ref{fig.Ld_w} is the noticeable decrease,
with the growth of $C$, of the penetration distance which the
projectile can achieve.  If one assumes that the dependence $n(z)$
satisfies the exponential-decay law (see (\ref{1.5})), then the curves
demonstrate that the dechanneling length is the decreasing function of
$C$.  This effect has been discuss in literature
(\cite{Biryukov,Taratin98}) in connection with the heavy particles
dechanneling in crystals bent with constant curvature.  Physically
clear quantitative arguments \cite{Biryukov}, based on the
relationship between the diffusion coefficient and the mean-square
angle of multiple scattering lead to the following estimate for the
dechanneling length \cite{Biryukov,laser} in the channel bent with the
mean curvature $1/\bar{R}$
\begin{equation}
\fl
L_d^e(\bar{R})
= 
\left(1 - R_c/\bar{R}\right)^2 \,
L_d^e(\infty),
\qquad
L_d^e(\infty)
=
{256 \over 9\pi^2}\,{ a_{\rm TF} \over r_{0} }\,
{ d \over L_c }\, \gamma
\label{stim9}
\end{equation}
here the quantity $R_c= \varepsilon/ U_{\rm max}^{\prime}$ is the
critical (minimal) radius consistent with the channelling condition in
a bent crystal (\ref{cond1}).  The quantity $L_c
=\ln\left(\sqrt{2\gamma}\, mc^2/I\right) - 23/24$, with $I = 16
Z^{0.9}$ eV standing for the Thomas-Fermi ionization potential of the
crystal atoms, is the Coulomb logarithm characterizing the ionization
losses of an ultra-relativistic positron in amorphous media with
account for the density effect (see e.g. \cite{Komarov,Sternheimer}).

The quantity $L_d^e(\infty)$ is the estimated value of the
dechanneling length for a straight channel.  The factor $\left(1 -
R_c/\bar{R}\right)^2$ is the correction due to the decrease of the
potential well of $U_{eff}(\rho)$ (see (\ref{2.25})) in bent crystal.

In periodically bent crystal the mean curvature radius is proportional
to $R_{\rm min}$ (in particular, for the shape (\ref{sinSz}),
$\bar{R}=\pi R_{\rm min}/2$), hence the ratio $R_c/\bar{R} \propto C$.
This estimate was used in \cite{laser,losses,new_lett} to
calculate the characteristics of the spontaneous and stimulated
emissions based on the AIR effect.

Basing on the estimate (\ref{stim9}) one deduces that the dechanneling
length is mainly determined by the value of the parameter $C$ and is
not influenced by the $a/d$ ratio.  This conclusion is supported by
the results of our calculations.  In figure (\ref{fig.Ld_si_all}) the
dependences $n(z)/n(0)$, obtained for different $C$ and $a/d$ values,
are compared.  It is seen that the variations between the curves with
different $a/d$ but the same $C$ is much smaller than for the curves
with the same $a/d$ ratio but different $C$ values.

The results of our calculations of the dechanneling lengths for $5$
GeV positrons in Si, Ge and W crystals obtained at different values of
the parameter $C$ but for fixed ratio $a/d=10$ are presented in table
\ref{Table1}. They are notated through $L_d^c$ (the quantities
$N_d^c=L_d^c/\lambda$ are the number of the undulator periods on the
scale of $L_d^c$). The presented values of $L_d^c$ were obtained by
fitting the smooth part of the curves $n(z)/n(0)$ (i.e. for the
distances $z$ beyond the region of the steep decrease) with the
exponent $\exp\left(-z/L_d^c\right)$.

\begin{table}
\caption{Dechanneling lengths for 5 GeV positron channeling
along the $(110)$ planes for various crystals and for various values
of the parameter $C$ (eq. (\protect\ref{cond1})).  The data correspond
to the shape function $S(z)=a\sin(2\pi z/\lambda)$.  The $a/d$ ratio
equals $10$ except for the case $C=0$ (the straight channel).  The
quantity $L_d^c$ presents the results of our calculations,
$N_d^c=L_d^c/\lambda$ is the corresponding number of the undulator
periods, $L_d^e$ is the dechanneling length estimated according to
(\protect\ref{stim9}), $N_d^e=L_d^e/\lambda$.  Other parameters are:
$d$ is the interplanar spacing, $R_c= \varepsilon/ U_{\rm
max}^{\prime}$ is the critical (minimal) radius consistent with the
condition (\protect\ref{cond1}), $\om$ is the energy of the first
harmonic of the AIR for the forward emission (see
(\protect\ref{1.4a})), $p$ is the undulator parameter,.}
\begin{indented}
\item[]\begin{tabular}{@{}rrrrrrrrr}
\br
Crystal:& Si,   &$d=$&$1.92$\AA,        &$R_c=$&$0.78$ cm&        &     &\\  
  $C$   & $\Rmin$ &$\lambda$  &$L_d^e$ & $L_d^c$ &$N_d^e$&$N_d^c$&$\om$&$p$\\
        &  cm     & $\mu$ m   &  cm    &  cm     &       &    &  MeV   & \\
 0.00   &$\infty$ &   -       &  0.312 &  0.463  &   -   &  - &   -  &-\\
 0.05   & 15.697  &  100.9   &   0.281 &  0.430  &   25  &39  &1.38& 1.08 \\
 0.10   &  7.849  &  77.1    &   0.253 &  0.393  &   32  &51  &1.42& 1.53 \\
 0.15   &  5.232  &  63.0    &   0.225 &  0.321  &   35  &51  &1.37& 1.87 \\
 0.20   &  3.924  &  54.5    &   0.200 &  0.223  &   36  &41  &1.31& 2.16 \\
 0.25   &  3.140  &  48.8    &   0.175 &  0.170  &   35  &35  &1.24& 2.42 \\
 0.30   &  2.616  &  44.5    &   0.153 &  0.102  &   34  &23  &1.18& 2.65 \\
 0.35   &  2.243  &  41.2    &   0.132 &  0.078  &   31  &19  &1.13& 2.86 \\
 0.40   &  1.962  &  38.6    &   0.112 &  0.042  &   29  &11  &1.09& 3.06 \\
\br
Crystal:& Ge,   &$d=$  & $2.00$\AA,     &$R_c=$&$0.42$ cm &        &     &\\
  $C$   & $\Rmin$ &$\lambda$  &$L_d^e$ & $L_d^c$ &$N_d^e$&$N_d^c$&$\om$&$p$\\
        &  cm     & $\mu$ m   &  cm    &  cm     &       &    &  MeV   & \\
 0.00   &$\infty$ &   -       &  0.263  & 0.513   &  -    &  -    &   -  &-\\
 0.05   & 8.465   & 81.8     &   0.237 & 0.450   &  29   &55  &1.37& 1.50\\ 
 0.10   & 4.232   & 57.8     &   0.213 & 0.364   &  36   &63  &1.26& 2.13\\
 0.15   & 2.822   & 47.2     &   0.190 & 0.269   &  40   &57  &1.15& 2.61\\
 0.20   & 2.116   & 40.9     &   0.168 & 0.176   &  41   &43  &1.05& 3.01\\
 0.25   & 1.693   & 36.6     &   0.148 & 0.095   &  40   &26  &0.98& 3.36\\
 0.30   & 1.411   & 33.4     &   0.129 & 0.060   &  38   &18  &0.92& 3.68\\
 0.35   & 1.209   & 30.9     &   0.111 & 0.028   &  35   & 9  &0.86& 3.98\\ 
 0.40   & 1.058   & 28.9     &   0.095 & 0.012   &  32   & 4  &0.82& 4.25\\
\br
Crystal:& W,    &$d=$  & $2.45$\AA,  &$R_c=$  &$0.10$ cm &       &      &\\  
  $C$   & $\Rmin$ &$\lambda$  &$L_d^e$ & $L_d^c$ &$N_d^e$&$N_d^c$&$\om$&$p$\\
        &  cm     & $\mu$ m   &  cm    &  cm     &       &    &  MeV   & \\
 0.00   &$\infty$ &   -       &  0.263  & 0.786   &  -    &  -    &   -  &-\\
 0.05   & 2.018   & 42.2     &  0.215  & 0.637   &  50   &151 &0.89& 3.26\\ 
 0.10   & 1.009   & 29.9     &  0.193  & 0.453   &  64   &152 &0.69& 4.61\\
 0.15   & 0.673   & 24.4     &  0.172  & 0.226   &  70   &93  &0.58& 5.64\\
 0.20   & 0.505   & 21.1     &  0.153  & 0.027   &  72   &13  &0.51& 6.52\\ 
 0.25   & 0.404   & 18.9     &  0.134  & 0.007   &  71   & 4  &0.46& 7.29\\
\br
\end{tabular}
\end{indented}
\label{Table1}
\end{table}

For each $C$ we included also the corresponding values of the
undulator period $\lambda$, the minimum curvature radius $R_{\rm min}=
1/(a k^2)=\lambda^2/4\pi^2 a$, the energy of the first harmonics of
AIR, $\hbar \omega_1$, obtained from (\ref{1.4a}) for the forward
emission ($\vartheta=0$), and the values of the undulator parameter
$p$.

The quantities $L_d^e$ represent the the dechanneling length estimated
from (\ref{stim9}), $N_d^e=L_d^e/\lambda$ is the corresponding number
of the undulator periods on the scale of $L_d^e$.

Comparing the calculated values $L_d^c$ with those estimated from
(\ref{stim9}) one notices that the latter underestimates the
dechanneling length in the case of small values of the parameter $C$:
$C<0.25$ for Si, $<0.20$ for Ge, and $<0.15$ for W.  It happens mainly
due to the fact that the quantity $L_d^e(\infty)$, which enters
(\ref{stim9}), was obtained by using the Lindhard approximation for
the interplanar potential $U(\rho)$.  The latter overestimates the
electron density inside the channel as compared with the Moli\`ere
approximation used in the present calculations.  Thus, the average
multiple scattering angle is higher if calculated within the Lindhard
approximation, and, consequently, the dechanneling lengths are lower.
With $C$ increasing the ratio $L_d^c/L_d^e$ becomes less than one.
This is explained as follows.  When the parameter $C$, and,
consequently, the curvature of the channel bending, increases, then
the minimum of the effective potential (\ref{2.25}) is shifted from
the midplane towards the atomic plane.  Hence, the channeled particle
moves in the region of higher electron density and, therefore, the
probability of the large-angle scattering increases
\cite{Biryukov,Taratin98}.  This is mechanism of the dechanneling,
which is additional to the bending dechanneling discussed above, is
not accounted for by the estimate (\ref{stim9}).  Comparing the
$L_d^c$ values for various $C$ it is seen that for large $C$ the
dechanneling length rapidly goes to zero, and the main reason for this
behaviour is the increase of the role of the large-angle scattering.

Another important feature to be mentioned is that the estimate
(\ref{stim9}) predicts the decrease in the dechanneling length for
heavier crystals, whereas the more accurate calculations within the
framework of the Moli\`ere approximation demonstrate that for
$C\in[0\dots 0.1]$ the largest dechanneling lengths are in W, followed
by the germanium crystal, and finally, the lowest values of $L_d^c$
are in the case of channeling in Si.  With $C$ increasing the
situations is reversed.  To explain this behaviour let us compare the
cases of Si and W crystals.  The maximum values of the interplanar
potential $U(\rho)$ at temperature $T=150$ K are equal to $22.9$ eV
for Si, and to $138.6$ eV for W.  Thus, the critical transverse energy
of the channeled positrons is approximately 6 times higher in W.
Hence, were the mean electron densities in these crystal equal then
the dechanneling length in W would be 6 times higher as compared to
the silicon case.  The ratio of the mean electron densities for the
inner regions of the $(110)$ channels in W and Si is approximately
$3.8$.  Therefore, the realistic estimate for the ratio
$\left(L_d\right)_{\rm W}/\left(L_d\right)_{\rm Si}$ is $6/3.8 \approx
1.6$ which is very close to that which can be obtained from table
\ref{Table1} for low $C$-values.  Our results for the relative
magnitudes of $L_d$ for a positron channeling in Si and W coincides
with the conclusion of made in \cite{Taratin95} where the calculations
of the dechanneling lengths were carried out for heavy particles in
the case of moderate bending of the crystals.

As $C$ increases, and the trajectory of the particle shifts towards
the atomic plane, the electron density in the region of the projectile
motion increases as well.  This increase is more pronounced for W (see
figure (\ref{fig.NeNr})) leading to the relation
$\left(L_d\right)_{\rm W}/\left(L_d\right)_{\rm Si} < 1$ in the range
$C> 0.15$.

The important conclusion which can be made on the basis of the results
presented in this section is the following.  To achieve higher yield
of the monochromatic radiation by means of the crystalline undulator
it is necessary to use heavier crystals and to restrict the parameter
$C$ by the range $0.05\dots 0.15$.  Indeed, for these $C$ values the
number of the undulator periods $N\approx N_d \gg 1$ (this is valid
for all crystals presented in the table), but the intensity of the
AIR, which is mainly defined by the factor $N^2 \approx N_d^2$ (see
(\ref{1.8}) and (\ref{2.17a})), will be almost by the order of
magnitude lager in the case of W than for Si and/or Ge crystals.
 
\section{Results of calculations of the spectral-angular and
spectral distributions of the AIR}
\label{specrtaAIR}

The simulation of the dechanneling process, described in the preceding
section, allows to establish the ranges of the parameter $N_d$ which
to a great extent defines the characteristics of the AIR radiation in
the presence of the dechanneling.  In this section we present the
results of calculations of the spectral-angular (see eq. (\ref{1.8}))
and the spectral distributions of the AIR.

The general features of the angular distribution of the radiation 
formed in an acoustically based undulator were discussed in
\cite{laser} where the effect of the dechanneling was taken into
account by restricting the length of the crystal (and, hence, the
number of the undulator periods) by the value $L_d^{e}$, which, as it
was demonstrated above, in many cases underestimates the dechanneling
length, and, consequently, leads to the underestimation of the photon
yield.

\begin{figure}
\hspace{2.5cm}\epsfig{file=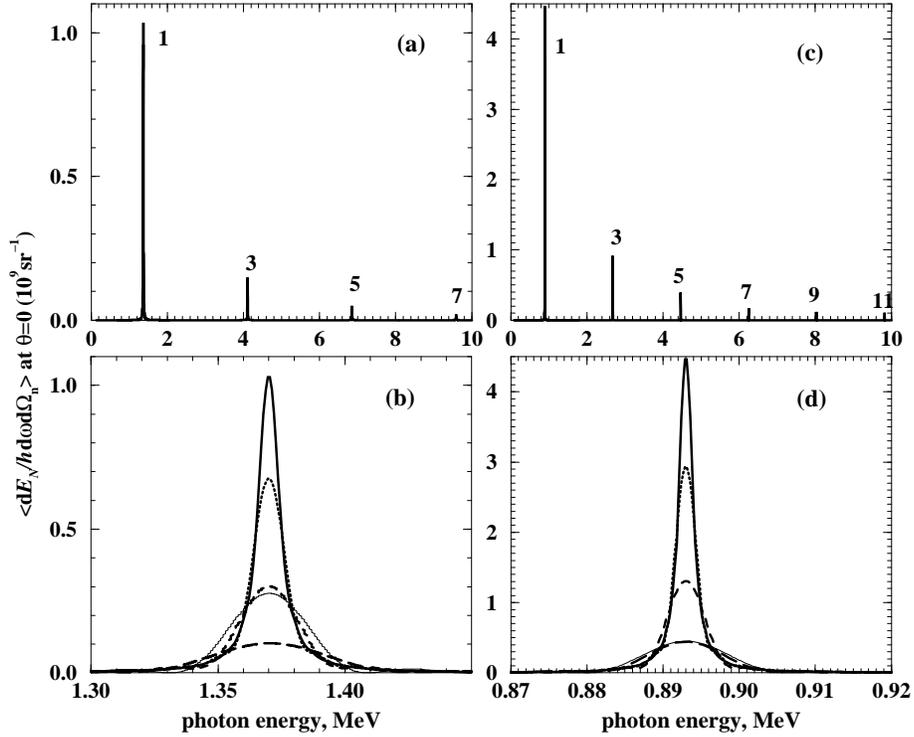,width=12cm, angle=0}
\caption{Spectral distribution (\protect\ref{1.8})  
(in $10^9\,{\rm sr}^{-1}$) at $\vartheta=0$ for 5 GeV positron
channeling along periodically bent $(110)$ planes in Si (figures (a)
and (b)) and W (figures (c) and (d)) crystals.  The parameters of the
crystals bending are as presented in the table for $C=0.15$ (for Si)
and $C=0.05$ (for W).  The upper figures (a) and (c) reproduce
$\left\langle {\d E_N/\hbar\, \d\omega\,\d\Omega_{\bf n}}\right\rangle
$ in the wide ranges of $\omega$ and correspond to $N=4\, N_d^{c}$.
The numbers enumerate the harmonics (in the case of the forward
emission the radiation occurs only in odd harmonics).  The profiles of
the first harmonic peak (figures (b) and (d)) are plotted for $N=4\,
N_d^{c}$ (solid lines), $N=2\, N_d^{c}$ (dotted lines), $N=N_d^{c}$
(dashed lines), $N=N_d^{c}/2$ (long-dashed lines).  In both figures
the thin solid line corresponds to the non-averaged spectrum
(\protect\ref{1.2}) calculated with the number undulator periods
$N=N_d^{e}$ which follows from (\protect\ref{stim9}).}
\label{fig.dsp}
\end{figure}

Figures \ref{fig.dsp}(a)--\ref{fig.dsp}(d) present the spectral
distribution (\ref{1.8}) of the radiation emitted along the undulator
axis, $\hbar^{-1}\langle \d E_N/\d\omega\,\d\Omega_{\bf
n}\rangle_{\vartheta=0^{\circ}}$, for 5 GeV positron channeling along
(110) planes in Si and W crystals.  The spectra were calculated by
using the parameters $\lambda$ and $p$ as indicated in table
\ref{Table1} in the lines $C=0.15$ for Si, and $C=0.05$ for W.  The
number of the undulator periods varied from $N=4\,N_d^c$, as in
figures \ref{fig.dsp}(a,c) and the solid curves in figures
\ref{fig.dsp}(b,d), to $N=N_d^c/2$ (the long-dashed lines in figures
\ref{fig.dsp}(b,d)). For fixed $\lambda$ different values of $N$ can
be achieved by changing the length of the crystal.  The sinusoidal
profile of the channel bending can be created by applying the
transverse acoustic wave.  The velocities of the transverse supersonic
waves propagating along (110) directions are equal $5.84$ cm/s for Si
\cite{SoundSi} and $5.17$ cm/s for W crystal \cite{Mason}.  Thus, the
corresponding frequencies of the acoustic wave needed to to obtain
$\lambda=63.0$ $\mu$m in Si and $\lambda=42.2$ $\mu$m in W (see the
table) are: $\nu_{\rm Si}=93$ MHz, $\nu_{\rm W}=123$ MHz.

The upper figures illustrate the structure of the spectral
distribution in wide ranges of the emitted photon energy. 
Each peak corresponds to the emission into the odd harmonics
\cite{Bazylev,Baier}, the energies of which follow from
(\ref{1.4a}) if putting $\vartheta=0^{\circ}$.  It is seen that all
harmonics are well separated: the distance $2 \hbar\omega_1$ between
two neighbouring peaks is $2.74$ MeV for Si and $1.78$ MeV in the case
of W, whilst the width of each peak $\hbar \Delta\omega$, estimated
from (\ref{2.17a}) (the case $N\gg N_d^c$), is $\approx 8.7$ keV for
Si and $\approx 2.5$ keV for W.  The intensity of the first-harmonic
peak in W is approximately 4.5 times larger than in the case of
silicon crystal.  This ratio is in agreement with the expression
(\ref{1.8}).  Indeed, the intensity of the first-harmonic peak is
proportional to the factor $p^2\, (\omega_1/\omega_0)^2\,\left\langle
D_N(1)\right\rangle$, which, in turn, is proportional to
$p^2/(2+p^2)^2\, \left(N_d^c\right)^2$ (see eqs. (\ref{1.4a}) and
(\ref{2.17a})).  Using the data presented in the table one obtains
that the ratio of these factors calculated for W and Si is $\approx
5$.

The difference in the magnitudes of the undulator parameters, $p=1.87$
for Si and $p=3.26$ for W, explains number of the harmonics visible in
the spectra \cite{Bazylev,Baier}.

Figures \ref{fig.dsp}(b,d) exhibit, in more detail, the structure of
the first-harmonic peaks.  For the sake of comparison we plotted the
curves corresponding to different values of the undulator periods.  It
is seen that for $N>N_d^c$ the intensity of the peaks is no longer
proportional to $N^2$, as it is in the case of the ideal undulator
without the dechanneling of the particles (Section \ref{Nodech}).  For
both Si and W crystals, the intensities of the radiation calculated at
$N\longrightarrow \infty$ exceeds those at $N=4\,N_d^c$ (the thick
solid curves in the figures) only by several per cent.  Thus, the
solid lines correspond to almost saturated intensities which are the
maximal ones for the used crystals, projectile energies and the
parameters of the crystalline undulator.  It is worth noting that
these intensities, obtained with realistic values of the dechanneling
lengths $L_d^c$, are noticeably higher than the ones corresponding to
$N>N_d^e$ (the thin solid lines in the figures). The latter were used
in \cite{laser} for the estimations of the yield of the AIR, both
spontaneous and stimulated.  It it seen from the figure that our
earlier estimations were adequate for light crystals (the maximum of
the thick solid curve exceeds that of the thin one by a factor of 3.5)
but they essentially underestimated the yield of the AIR in heavy
crystals, where the ratio of the intensities is almost 10.

\begin{figure}
\hspace{2.5cm}\epsfig{file=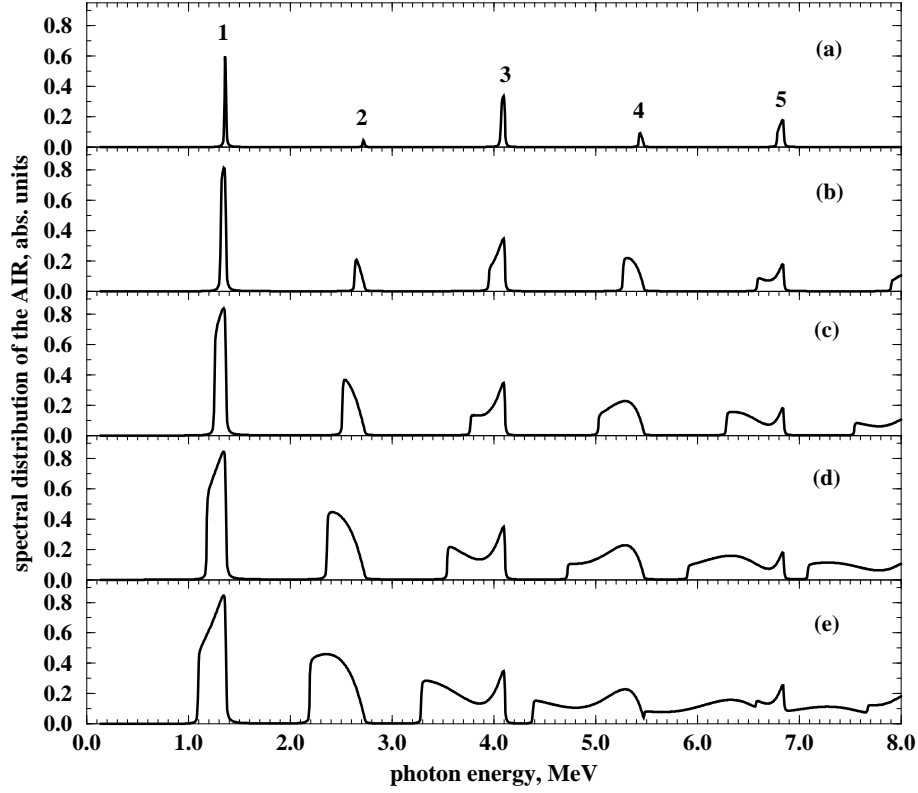,width=12cm, angle=0}
\caption{Spectral distribution (\protect\ref{sp.eq})  
for 5 GeV positron channeling
along periodically bent $(110)$ planes in 
Si. The five graphs correspond to different values of 
$\vartheta_{\rm max}$:
(a) $\vartheta_{\rm max}=0.1\, \vartheta_{\rm c}$,
(b) $\vartheta_{\rm max}=0.2\, \vartheta_{\rm c}$,
(c) $\vartheta_{\rm max}=0.3\, \vartheta_{\rm c}$,
(d) $\vartheta_{\rm max}=0.4\, \vartheta_{\rm c}$,
(e) $\vartheta_{\rm max}=0.5\, \vartheta_{\rm c}$,
with $\vartheta_{\rm c}=p/\gamma=0.187$ $m$rad.
The numbers in figure (a) enumerate the harmonics.}
\label{fig.sp_si}
\end{figure}

\begin{figure}
\hspace{2.5cm}\epsfig{file=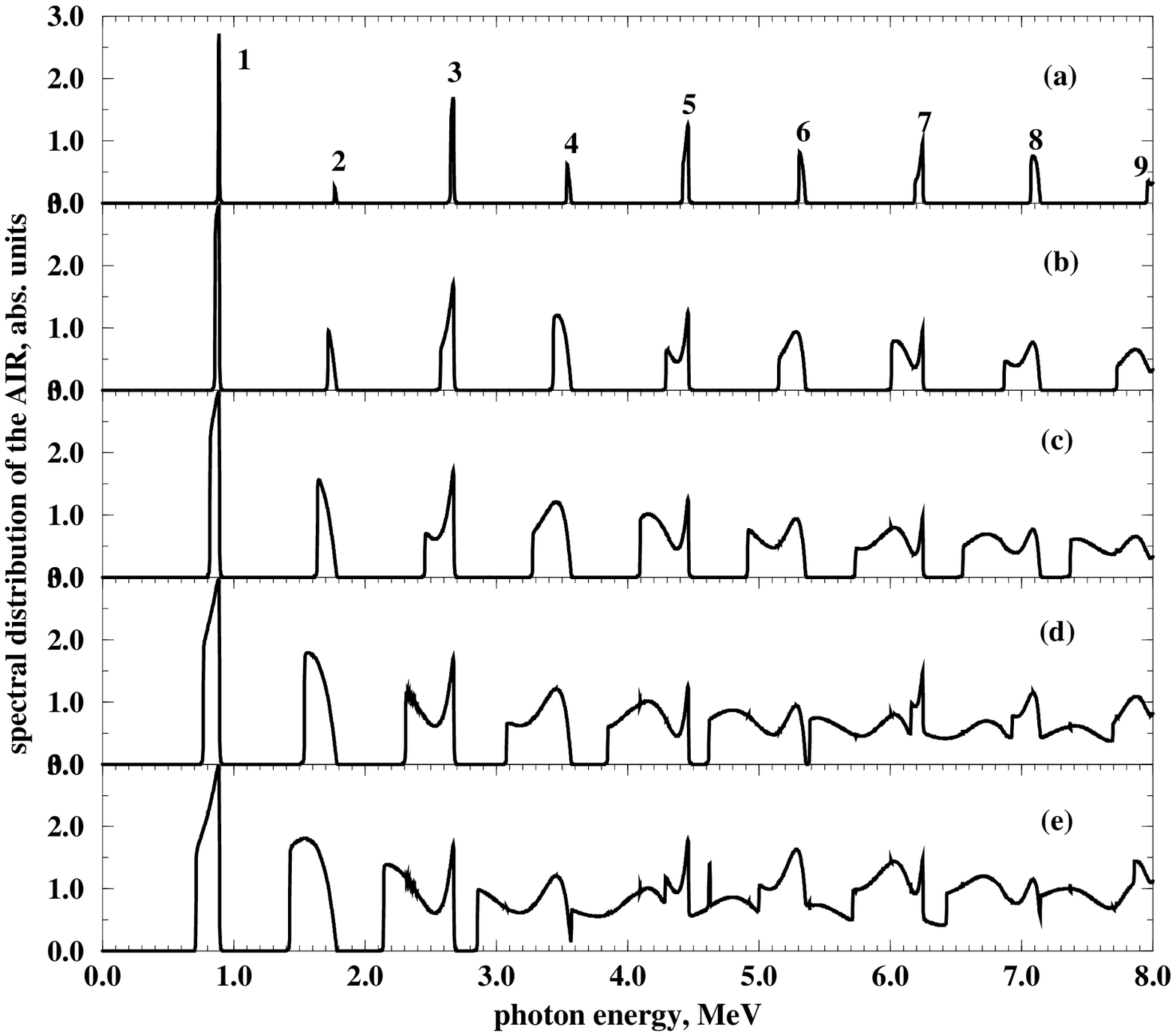,width=12cm, angle=0}
\caption{Same as in figure (\protect\ref{fig.sp_si})  
but for tungsten crystal.
The parameter
$\vartheta_{\rm c}=p/\gamma$ equals to $0.187$ $m$rad.}
\label{fig.sp_w}
\end{figure}

In figures \ref{fig.sp_si} and \ref{fig.sp_w} we present the results
of the spectral distribution of the radiation $\hbar^{-1}\langle \d
E_N/\d\omega\rangle$ obtained by integration of the r.h.s. of
(\ref{1.8}) over the emission angle $\d\Omega_{\bf n}$.  The
calculations were performed for 5 GeV positrons channeled in Si and W
with the bending parameters as those mentioned above.

It is known from general theory of the undulator radiation by an
ultra-relativistic projectile \cite{Bazylev,Baier} that it is emitted
into the narrow cone along the undulator axis. The opening angle of
this cone, $\vartheta_{\rm c}$ depends on the two parameters, namely,
on the relativistic factor $\gamma$ and on the undulator parameter
$p$.  For $p^2 \ll 1$ (the so-called dipole case) the radiation is
concentrated within the cone with the opening angle $\sim 1/\gamma$.
In the opposite limit, $p^2
\gg 1$ (the non-dipole case) the opening angle becomes larger, $\sim
p/\gamma$.  Uniting both limiting cases one can state that for given
$\gamma$ and $p$ the maximum angle of the emission can be estimated as
$\vartheta_{\rm c} \sim {\rm max}\left\{\gamma^{-1}, p
\gamma^{-1}\right\} \ll 1$.  In the case of a planar undulator, which
is considered here, and for $p>1$ the cone is asymmetric with respect
to the azimuthal ($\varphi$) angle of the photon emission
\cite{laser}.  The opening angle is the largest for the emission
within the undulator plane ($\varphi=0^{\circ}$) and reaches its
minimum value for the emission into the perpendicular plane
($\varphi=90^{\circ}$).

The five graphs in each of the figures reproduce the dependence of the 
quantity
\begin{eqnarray}
\left\langle 
{\d E_N \over \d\omega}\right\rangle_{\vartheta\leq \vartheta_{\rm max}} 
=
\int_0^{\vartheta_{\rm max}}\vartheta \d \vartheta \,
\int_0^{2\pi}\d \varphi \,
\left\langle 
{\d E_N \over \d\omega\,\d\Omega_{\bf n}}\right\rangle 
\label{sp.eq}
\end{eqnarray}
on $\omega$ for different values of the parameter $\vartheta_{\rm
max}$ as it is indicated in the figures.

The parameter $\vartheta_{\rm c}$, defined above, equals to $0.187$
$m$rad for Si and $0.326$ $m$rad for W crystals.

The graphs (a)-(e) in each figure illustrate how the pattern of the
spectral distribution changes with enlarging the opening cone
$\vartheta_{\rm max}$.  For small $\vartheta_{\rm max}$, when only the
radiation emitted in the nearly forward direction is taken into
account (the graphs (a) in the figures), the shape of the distribution
(\ref{sp.eq}) is close to that exhibited in figures \ref{fig.dsp} (a)
and (c).  The radiation into the odd harmonics dominates over the
even-harmonics peaks which, nevertheless, become visible in contrast
with the case of emission at $\vartheta=0$.

\begin{figure}
\hspace{2.5cm}\epsfig{file=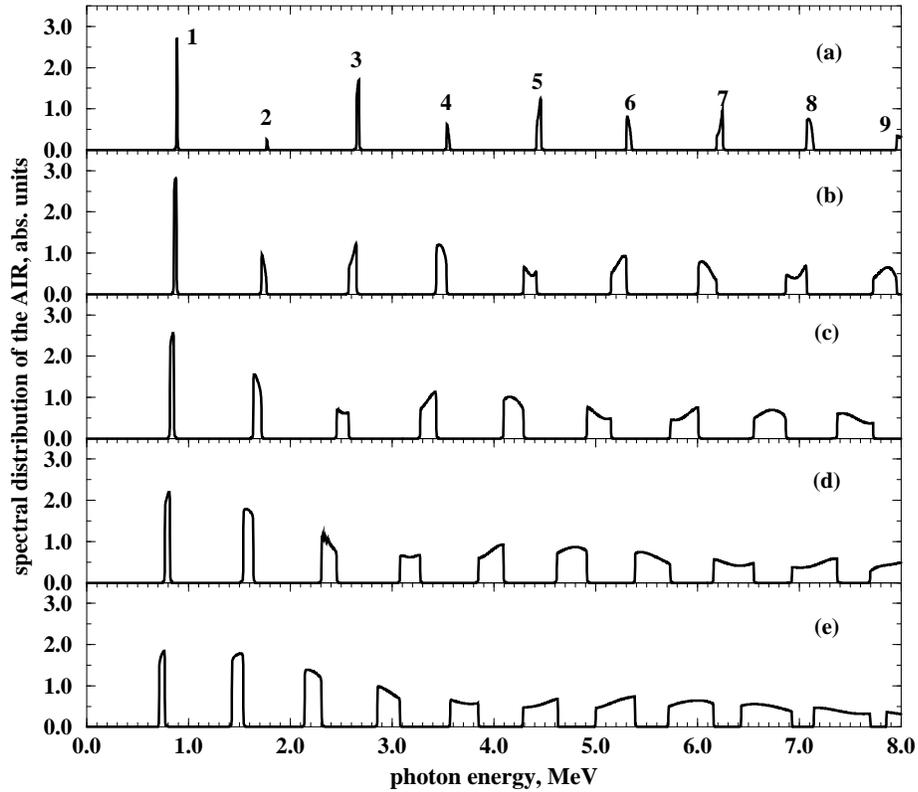,width=12cm, angle=0}
\caption{Spectral distribution of the AIR for different intervals of
the emission angle $\vartheta$:
(a) $0 \leq  \vartheta\leq 0.1\, \vartheta_{\rm c}$,
(b) $0.1\, \vartheta_{\rm c}\leq  \vartheta \leq 0.2\, \vartheta_{\rm c}$,
(c) $0.2\, \vartheta_{\rm c}\leq  \vartheta \leq 0.3\, \vartheta_{\rm c}$,
(d) $0.3\, \vartheta_{\rm c}\leq  \vartheta \leq 0.4\, \vartheta_{\rm c}$,
(e) $0.4\, \vartheta_{\rm c}\leq  \vartheta \leq 0.5\, \vartheta_{\rm c}$,
with $\vartheta_{\rm c}=0.187$ $m$rad.
The graphs correspond  to 5 GeV positron channeling
along periodically bent $(110)$ planes in W. 
The numbers in figure (a) enumerate the harmonics.}
\label{fig.diffy}
\end{figure}

As the cone of emission becomes larger (the graphs (c)-(e) in the
figures) the width of the peaks grows and their shape becomes
asymmetric.  The enhancement of the width follows from
eq. (\ref{1.4a}) which connects the harmonics frequencies with the
emission angle, and from figure \ref{fig.DN}.  For a given number $k$
of the harmonic its frequency $\omega_k(\vartheta)$ is the decreasing
function of the emission angle, so that the center of the peak shifts
towards lower values of the frequency as $\vartheta$ grows.  Hence,
the integration of (\ref{1.8}) over the interval $\vartheta,
\vartheta+\Delta\vartheta$ leads to the appearing of the emission
within the frequency range
$\sim\left[\omega_k(\vartheta)-\Delta\omega,\
\omega_k(\vartheta)\right]$ where
\begin{eqnarray*}
\Delta\omega =
\omega_k(\vartheta)\,
{4\gamma^2\, \vartheta\,\Delta\vartheta 
\over
2+p^2 + 2\gamma^2\vartheta}.
\end{eqnarray*}
We illustrate this estimate by figure \ref{fig.diffy}, where the
graphs (a)-(e) correspond to the contributions of different intervals
of the emission angle to the spectral distribution of the AIR for 5
GeV positron in tungsten.  It is clearly seen that each interval of
$\theta$ corresponds to the emission radiated within particular ranges
of $\omega$.

In figures \ref{fig.sp_si} and \ref{fig.sp_w} the graphs (e)
correspond to the value $\vartheta_{\rm max}=0.5\, \vartheta_{\rm c}$
for the upper limit of integration over the emission angles.  It is
seen that initially well-separated narrow peaks (the graphs (a)) have
merged, except for the first three harmonics.  The further increase of
$\vartheta_{\rm max}$ up to $\vartheta_{\rm c}$ produces the
continuous spectrum of radiation although with the irregularities in
the vicinities of $\omega_k(0)$.

\section{Conclusions}

In this work we have described the general formalism and the algorithm
for effective calculations of the characteristics of the AIR as well
as of the dechanneling length of positrons in periodically bent
crystals.

The calculations performed in the present paper confirm the main
result of our previous considerations concerning the possibility to
create a powerful and easily tunable source of monochromatic radiation
in the $X$ or/and $\gamma$ range by means of crystalline undulator.
They also demonstrate that the knowledge of the accurate values of the
dechanneling lengths for various projectile energies, crystals and
crystallographic palnes, and various parameters of the profile
function $S(z)$ is essential for obtaining reliable data on the
intensity of the AIR.

The numeric results presented above refer to a particular shape of the
bent channel which can be achieved by applying monochromatic
transverse supersonic wave.  However, both the formalism and the
computer code allows to investigate, by comparatively simple means,
the radiative spectra formed during the channeling of
ultra-relativistic positrons along the crystallographic planes whose
shape is described by arbitrary periodic shape function.

In this connection we want to mention the problem concerning the
possibility to generate the AIR photons not in the range of up to
several MeV but much higher.  It can be achieved by using positron
beams of higher energies, $\varepsilon =10 \dots 100$ GeV.  The
increase of $\varepsilon$ results in the increase of both the emitted
photon energies and the dechanneling length.  The latter leads to the
increase in the intensity of the AIR.  The obvious difficulty of the
stable work of the crystalline undulator in this range of
$\varepsilon$ is in sharp increase of the radiative energy losses
which are mainly due the ordinary channeling radiation.  The radiative
losses lead to a noticeable decrease of the projectile energy during
the passage through the undulator.  Hence, the use of the shape
function with constant parameters $a$ and $\lambda$ does not result in
the intensive monochromatic undulator radiation.  However, we believe,
that this problem can be overcome by considering $S(z)$ with the
varied parameters, $a(z)$ and $\lambda(z)$.  Then, provided the
dependence $\varepsilon(z)$ ($z=c t$) is known it is possible to
adjust the parameters $a(z)$, $\lambda(z)$ so that the characteristic
frequencies of the AIR will not change as the projectile penetrates
through the crystal, and, therefore, the monochromaticity will be
restored.  This work is in progress at the moment and will become the
subject of another publication in the near future.

\ack

The research was supported by DFG, GSI, and BMBF.  AVK and AVS
acknowledge the support from the Alexander von Humboldt Foundation.


\end{document}